\documentclass[twocolumn,showpacs,showkeys,footinbib,superscriptaddress]{revtex4}
\usepackage{graphicx}
\usepackage{amsmath,amsfonts,amssymb,bm,color}

\begin{document}

\title{Nuclear fusion in muonic deuterium-helium complex}

\author{V.M.~Bystritsky}
\altaffiliation{Corresponding author}
\email{bystvm@nusun.jinr.ru}
\affiliation{Joint Institute for Nuclear Research, Dubna 141980,
Russia}

\author{M.~Filipowicz}
\affiliation{University of Science and Technology, Fac.~of Fuels and
Energy, PL--30059 Cracow, Poland}

\author{V.V.~Gerasimov}
\affiliation{Joint Institute for Nuclear Research, Dubna 141980,
Russia}

\author{P.E.~Knowles}
\affiliation{Department of Physics, University of Fribourg, CH--1700
Fribourg, Switzerland}

\author{F.~Mulhauser}
\affiliation{University of Illinois at
  Urbana--Champaign, Urbana, Illinois 61801, USA}

\author{N.P.~Popov}
\altaffiliation{Visiting Professor}
\affiliation{University of Science and Technology,
Fac.~Phys.~Nucl.~Techniques, PL--30059 Cracow, Poland}

\author{V.A.~Stolupin}
\affiliation{Joint Institute for Nuclear Research, Dubna 141980,
Russia}

\author{V.P.~Volnykh}
\affiliation{Joint Institute for Nuclear Research, Dubna 141980,
Russia}

\author{J.~Wo\'zniak}
\affiliation{University of Science and Technology,
Fac.~Phys.~Nucl.~Techniques, PL--30059 Cracow, Poland}


\begin{abstract}
Experimental study of the nuclear fusion reaction in
charge-asymmetrical $d \mu{}^{3}\mathrm{He}$ complex
($d \mu{}^{3}\mathrm{He} \rightarrow \alpha$ (3.5~MeV) + $p$
(14.64~MeV)) is presented.
The 14.64~MeV protons were detected by three pairs of Si($dE-E$)
telescopes placed around the cryogenic target filled with the
$\mathrm{D}_2 + {}^{3}\mathrm{He}$ gas at 34~K\@.
The 6.85~keV $\gamma$~rays emitted during the de-excitation of the $d
\mu{}^3 \mathrm{He}$ complex were detected by a germanium detector.
The measurements were performed at two $\mathrm{D}_2 +
^{3}\mathrm{He}$ target densities, $\varphi = 0.0585$ and $\varphi =
0.169$ (relative to liquid hydrogen density) with an atomic
concentration of $^{3}\mathrm{He}$ $c_{{}^3\mathrm{He}} = 0.0469$.
The values of the effective rate of nuclear fusion in $d
\mu{}^3\mathrm{He}$ was obtained for the first time:
$\tilde{\lambda}_{f} = (4.5^{+2.6}_{-2.0}) \times 10^5 \,
\mbox{s}^{-1} (\varphi = 0.0585)$; $\tilde{\lambda}_{f} =
(6.9^{+3.6}_{-3.0}) \times 10^5 \, \mbox{s}^{-1} (\varphi = 0.168)$.
The $J=0$ nuclear fusion rate in $d \mu{}^3\mathrm{He}$ was derived:
$\lambda^{J=0}_{f} = (9.7^{+5.7}_{-2.6}) \times 10^5 \, \mbox{s}^{-1}\
(\varphi = 0.0585)$; $ \lambda^{J=0}_{f}=(12.4^{+6.5}_{-5.4}) \times 10^5
\, \mbox{s}^{-1}\ (\varphi = 0.168)$.
\end{abstract}


\maketitle


\section{Introduction}
\label{sec:introduction}

\begin{figure*}[t]
\centerline{
\setlength{\unitlength}{0.75mm} \thicklines
\begin{picture}(200,145)(0,-10)
\put(10,85){\oval(10,10)} \put(10,85){\makebox(0,0){$\mu^{-}$}}
\put(15,85){\vector(1,0){8}} \put(20,90){\makebox(0,0){$W_{d}$}}
\put(12,80){\vector(1,-1){15}}
\put(11,70){\makebox(0,0){$(1\!-\!W_{d})$}}
\put(35,90){\vector(0,1){15}} \put(30,97){\makebox(0,0){$q_{1s}$}}
\put(35,80){\vector(0,-1){15}}
\put(46,73){\makebox(0,0){$(1-q_{1s})$}}
\put(33,85){\oval(20,10)} \put(35,85){\makebox(0,0){$(\mu d
)^{*}$}}
\put(33,110){\oval(20,10)} \put(33,110){\makebox(0,0){$(\mu
d)_{1s,F}$}}
\put(33,60){\oval(20,10)}
\put(33,60){\makebox(0,0){$\mu{}^3\mathrm{He} $}}
\put(35,55){\vector(0,-1){10}}
\put(28,50){\makebox(0,0){$\lambda_\mathrm{cap}^{\mathrm{He}}$}}
\put(6,22){\framebox(24,10){}} \put(18,27){\makebox(0,0){$p + 2 n
+ \nu_\mu$}} \put(40,22){\framebox(24,10){}}
\put(52,27){\makebox(0,0){$d + n + \nu_\mu$}}
\put(25,0){\framebox(20,10){}} \put(35,5){\makebox(0,0){$t +
\nu_\mu$}}
\put(35,45){\vector(1,-1){13}}
\put(48,40){\makebox(0,0){$\lambda_\mathrm{cap}^d$}}
\put(35,45){\vector(-1,-1){13}}
\put(22,40){\makebox(0,0){$\lambda_\mathrm{cap}^p$}}
\put(35,45){\vector(0,-1){35}}
\put(42,16){\makebox(0,0){$\lambda_\mathrm{cap}^t$}}
\put(43,110){\line(1,0){7}} \put(58,110){\makebox(0,0){$\varphi \,
\mathrm {\mathrm c}_{d} \tilde \lambda_F$}}
\put(66,110){\vector(1,0){7}}
\put(40,105){\line(2,-3){28}} \put(70,60){\makebox(0,0){$\varphi
\, \mathrm {\mathrm c}_{{}^3\mathrm{He}}
\lambda_{d{}^3\mathrm{He}}$}} \put(72,57){\vector(2,-3){8}}
\put(68,15){\framebox(42,30){}}
\put(89,35){\makebox(0,0){$[(d\mu{}^3\mathrm{He})^* e^-]^+ +
e^-$}} \put(78,25){\makebox(0,0){2p$\sigma$}}
\put(100,25){\makebox(0,0){J=1}}
\put(150,15){\framebox(42,30){}}
\put(171,35){\makebox(0,0){$[(d\mu{}^3\mathrm{He})^* e^-]^+ +
e^-$}} \put(160,25){\makebox(0,0){2p$\sigma$}}
\put(182,25){\makebox(0,0){J=0}}
\put(110,30){\line(1,0){15}}
\put(132,30){\makebox(0,0){$\tilde\lambda_{10}$}}
\put(138,30){\vector(1,0){12}}
\put(100,60){\framebox(70,30){}}
\put(135,80){\makebox(0,0){$\alpha + \mu ^{-} + p
\:(14.64\:\mathrm{MeV})$}}
\put(135,70){\makebox(0,0){$\mu{}^{5}\!\mathrm{Li}+\gamma
\:(16.4\:\mathrm{MeV})$}}
\put(115,0){\framebox(30,16){}} \put(130,8){\makebox(0,0){$[\mu
{}^3\mathrm{He} + d]$}}
\put(90,15){\line(0,-1){5}} \put(90,10){\vector(1,0){25}}
\put(170,15){\line(0,-1){5}} \put(170,10){\vector(-1,0){25}}
\put(80,2){\makebox(0,0){$\lambda^{J=1}_{p}$,}}
\put(93,2){\makebox(0,0){$\lambda^{J=1}_{e}$,}}
\put(107,2){\makebox(0,0){$\lambda^{J=1}_{\gamma}$}}
\put(153,2){\makebox(0,0){$\lambda^{J=0}_{p}$,}}
\put(166,2){\makebox(0,0){$\lambda^{J=0}_{e}$,}}
\put(179,2){\makebox(0,0){$\lambda^{J=0}_{\gamma}$}}
\put(90,45){\line(0,1){25}} \put(90,70){\vector(1,0){10}}
\put(180,45){\line(0,1){25}} \put(180,70){\vector(-1,0){10}}
\put(93,75){\makebox(0,0){$\lambda^{J=1}_f$}}
\put(180,75){\makebox(0,0){$\lambda^{J=0}_f$}}
\put(83,110){\oval(20,10)} \put(83,110){\makebox(0,0){$d \mu d $}}
\put(93,110){\vector(1,0){17}}
\put(110,100){\line(0,1){22}}
\put(110,122){\vector(1,0){15}}
\put(117,125){\makebox(0,0){$\beta_F$}}
\put(125,116){\line(0,1){12}}
\put(125,128){\vector(1,0){15}}
\put(133,125){\makebox(0,0){$\omega_{d}$}}
\put(150,128){\oval(20,10)} \put(150,128){\makebox(0,0){$\mu
{}^3\mathrm{He}$}} \put(165,128){\makebox(0,0){$+ n$}}
\put(125,116){\vector(1,0){15}}
\put(131,119){\makebox(0,0){$1\!-\!\omega_{d}$}}
\put(170,116){\makebox(0,0){$\:{}^3\mathrm{He} + \mu^{-} +
n\:{\mathrm{(2.45~MeV)}}$}}
\put(110,100){\vector(1,0){30}}
\put(120,103){\makebox(0,0){$1-\beta_F$}}
\put(168,100){\makebox(0,0){$t + \mu^{-} + p\:{\rm (3.02~MeV)}$}}
\put(150,133){\line(0,1){2}} \put(150,135){\line(-1,0){150}}
\put(0,135){\vector(0,-1){40}} \put(0,95){\line(0,-1){35}}
\put(0,60){\vector(1,0){23}}
\put(130,0){\line(0,-1){3}} \put(130,-3){\line(-1,0){130}}
\put(0,37){\line(0,1){23}} \put(0,-3){\vector(0,1){40}}
\end{picture}
}
                \caption{Scheme of mu-atomic and mu-molecular
                processes occurring at stops of negative muons in the
                $\mathrm{D}_2 +{}^{3}\mathrm{He}$ mixture.}
\label{fig:scheme}
\end{figure*}
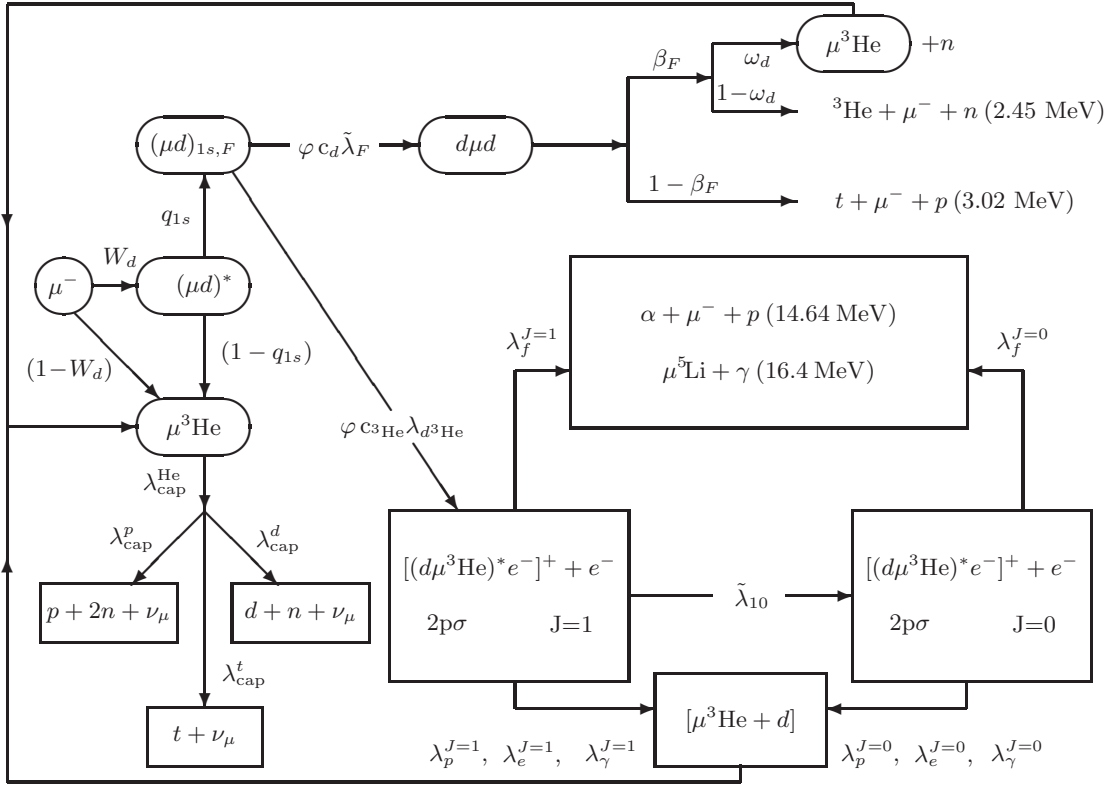

The formation of muonic molecules of hydrogen isotopes and their
nuclear reactions have been the subject of many experimental and
theoretical
studies~\cite{marsh01,petit01,petit92,cohen90,ponom90,ponom01,nagam01,bogda90c}.
As to the studies of formation of charge-asymmetrical muonic molecules
like $h \mu Z$ ($h = p, d, t$, $Z$ are nuclei with a charge $Z>1$) and
their respective nuclear fusion, the situation slightly different.
What gave an impetus to study such systems was the theoretical
prediction and experimental observation of the molecular mechanism for
charge exchange (MMCE) of $p \mu$ atoms on He
nuclei~\cite{arist81,bystr83b}.
Essentially, the mechanism is reduced to the following.
Colliding with a He atom in a $H$--He mixture ($H = \mathrm{H}_2,
\mathrm{D}_2, \mathrm{T}_2$ and $\mathrm{He} = {}^{3}\mathrm{He},
{}^{4}\mathrm{He}$), the muonic hydrogen atom forms a muonic complex
$h\mu \mathrm{He}$ in the excited $2p\sigma$ state.
In the case of a deuterium--helium mixture, the complex may then
decays from this state (see Fig.~\ref{fig:scheme}) via one of three
channels
\newcounter{bean}
\setcounter{bean}{0}
\renewcommand{\theequation}{1\alph{bean}}
\begin{alignat}{2}
\label{eq1}
d \mu + {\mathrm{He}} \stackrel{\lambda_{d
\mathrm{He}}}{\longrightarrow} &\ [(d \mu {\mathrm{He}} )^* e^-]^+
& +\ e^- \mbox{ \hskip .5in\ } \nonumber \\
& \hspace{0.6cm} \downarrow &  \nonumber \\
& \ \ \ \ \stackrel{\lambda _{\gamma}}{\longrightarrow} & [(d \mu
{\mathrm{He}})^+ e^-] +\gamma \addtocounter{bean}{1} \\
& \ \ \ \ \stackrel{\lambda_p}{\longrightarrow} & [(\mu
{\mathrm{He}})^+_{1s} e^-] + d \addtocounter{bean}{1} \\
& \ \ \ \ \stackrel{\lambda_e}{\longrightarrow} & (\mu
{\mathrm{He}})^+_{1s} + d + e^- \, .  \addtocounter{bean}{1}
\end{alignat}
If $\mathrm{He} = {}^{3}\mathrm{He}$, fusion reactions may occur
\setcounter{bean}{0}
\renewcommand{\theequation}{2\alph{bean}}
\begin{eqnarray}
  \label{eq2}
  d \mu{}^3{\mathrm{He}} &
   \stackrel{\tilde{\lambda}_f}{\longrightarrow} & \alpha + \mu +
   \mathrm{p\ (14.64\ MeV)} \addtocounter{bean}{1} \\
  & \stackrel{\tilde{\lambda}_{f\Gamma}}{\longrightarrow} &
  \mu^5\mathrm{Li} + \gamma\ \mathrm{(16.4\ MeV)}\, .
  \addtocounter{bean}{1}
\end{eqnarray}
Thus, the fusion proceeds by the formation of a $d \mu$ atom, which,
when incident on a ${}^{3}\mathrm{He}$ atom, forms the $d\mu
{}^{3}\mathrm{He}$ molecular system.
This molecule has two primary spin states, $J=1$ and
$J=0$~\footnote{$J$ denotes the total angular momentum of the three
particles.}; formation favors the former, fusion the
latter~\cite{bogda98}.
In Eqs.~(\ref{eq1}--c), $\lambda _{\gamma}$ is the $(d\mu
\mathrm{He})^*$ molecular decay channel for the 6.85~keV $\gamma$--ray
emission, $\lambda_e$ for the Auger decay, and $\lambda_p$ for the
break--up process.
The $d\mu \mathrm{He}$ molecule is formed with a rate $\lambda_{d
\mathrm{He}}$.
The main fusion process, Eq.~(\ref{eq2}a), occurs with the rate
$\tilde{\lambda}_f$, whereas the reaction~(\ref{eq2}b), with the
associated rate $\tilde{\lambda}_{f\Gamma}$ has a branching ratio on
the order of $10^{-(4,5)}$~\cite{cecil85}.

Interests in further study of charge-asymmetrical systems was caused
by first getting information on characteristics of the strong
interaction in the region of ultralow energies.
Secondly, it allows us to test the problem of three bodies interacting
via the Coulomb law.
More precisely, these studies may allow us to
\begin{itemize}
\item[--] check fundamental symmetries and to measure the main
characteristics of the strong interaction in the region of
astrophysical particle collision energies ($\sim$keV) in the entrance
channel.
It should be mentioned that nuclear fusion reactions in
charge-asymmetrical muonic molecules are characterized by the same
astrophysical range of energies~\cite{friar91}.
\item[--] test the calculation algorithm for rates of nuclear fusion
reactions in $\mu$-molecular complexes as well as for partial rates of
decay of these asymmetrical complexes via various channels.
\item[--] solve some existing astrophysical problems.
\end{itemize}
By now the experimental discovery of the MMCE has been confirmed in a
number of experiments on study of muon transfer from $h\mu$ to the He
isotopes.

Formation rates of the charge-asymmetrical $d \mu \mathrm{He}$, and $p
\mu \mathrm{He}$ systems were
measured~\cite{balin85,vonar89,bystr93d,bystr90,bystr90e,bystr90f,tresc98,gartn00,
bystr95d,bystr04b} and
calculated~\cite{ivano86,kravt86b,kinox93,gersh93,korob93b,czapl97b,belya95c,
czapl96c,belya97} with quite a good accuracy, and partial decay rates
of such complexes were found.

\renewcommand{\theequation}{\arabic{equation}}
\addtocounter{equation}{-3}

\begin{table}[ht]
\begin{ruledtabular}
      \caption{Experimental and calculated nuclear fusion rates, in
               s$^{-1}$, in the $d\mu {}^{3}\mathrm{He}$
               complex. $\tilde{\lambda}_{f}$ is the effective rate of
               fusion reaction~(\ref{eq2}a), ${\lambda}_{f}^{J=0}$ and
               ${\lambda}_{f}^{J=1}$ are the rates of fusion
               reaction~(\ref{eq2}a) in the $d\mu {}^3\mathrm{He}$
               complex in the $J = 0$ and $J = 1$ states,
               respectively.}
\label{tab:rates}
\begin{tabular}{ccccccccccccc}
& \multicolumn{12}{c}{experiment} \\
Refs. & \multicolumn{3}{c}{\cite{balin92b}} &
\multicolumn{3}{c}{\cite{balin98}}
& \multicolumn{3}{c}{\cite{maevx99}} &
  \multicolumn{3}{c}{\cite{delro99}} \\ \hline
$\tilde{\lambda}f$ &  \multicolumn{3}{c}{$\le 7 \times 10^7$} &
\multicolumn{3}{c}{$\le 1.6 \times 10^5$} &
\multicolumn{3}{c}{$\le 6 \times 10^4$}
&  \multicolumn{3}{c}{$\le 5 \times 10^5$} \\
\hline
& \multicolumn{12}{c}{theory}\\
Refs. & \multicolumn{2}{c}{\cite{kinox93}} &
\multicolumn{2}{c}{\cite{nagam89}}
&\multicolumn{2}{c}{\cite{penko97}} &
\multicolumn{2}{c}{\cite{czapl96,czapl98}} &
\multicolumn{2}{c}{\cite{harle89}} &
\multicolumn{2}{c}{\cite{bogda99}} \\ \hline
${\lambda}_{f}^{J=0} $ &
&&\multicolumn{2}{c}{$3 \times 10^8$} & \multicolumn{2}{c}{$3.8 \times
  10^6$}
& \multicolumn{2}{c}{$\sim 10^6$} & \multicolumn{2}{c}{$10^{11}$} &
\multicolumn{2}{c}{$1.9 \times 10^5$} \\
${\lambda}_{f}^{J=1}$ & \multicolumn{2}{c}{$10^6$} &
\multicolumn{8}{c}{ }& \multicolumn{2}{c}{$6.5 \times 10^2$} \\
\end{tabular}
\end{ruledtabular}
\end{table}

In the past five years interest in studying charge-asymmetrical
complexes and in particular fusion in the $d\mu {}^{3}\mathrm{He}$
system has revived.
Table~\ref{tab:rates} presents the calculated fusion rates of
deuterium and ${}^{3}\mathrm{He}$ nuclei in the $d\mu
{}^{3}\mathrm{He}$ complex in its states with the orbital momenta $J =
0$ and $J = 1$ and the experimental upper limits of the effective
fusion rate, $\tilde \lambda_f$, in the molecule averaged over the
populations of fine-structure states of the $d\mu {}^{3}\mathrm{He}$
complex.

The experimental study of nuclear fusion in the $d\mu
{}^{3}\mathrm{He}$ molecule is quite justified as far as detection of
the process is concerned because there might exist an intermediate
resonant compound state $^5\mathrm{Li}^*$ leading to the expected high
fusion rate which results from quite a large value of the $S$--factor
for the $d {}^{3}\mathrm{He}$ reaction~\cite{caugh88}.
However, as follows from the calculations presented in
Table~\ref{tab:rates}, the theoretical predictions of the fusion rate
in this molecule show a wide spread in value from $\sim 10^5 \,
\mbox{s}^{-1}$ to $10^{11} \, \mbox{s}^{-1}$.

The nuclear fusion rate in muonic molecules is usually calculated on
the basis of Jackson's idea~\cite{jacks57} which allows the
factorization of nuclear and molecular coordinates.
In this case the nuclear fusion rate $\lambda_{nf}$ is given by
\begin{equation}
\label{eq3}
\lambda_{nf} = \frac{S}{(\pi M Z_1 Z_2)} \times \left| \Psi_{sc}(0)
\right|^2 ,
\end{equation}
which is defined by the astrophysical $S$--factor, the reduced mass of
the system $M$, the charges of nuclei in the muonic molecule $Z_1$ and
$Z_2$, and the three--body system wave function $\Psi_{sc}(0)$
averaged over the muon degrees of freedom and taken at distances
comparable with the size of the nuclei, i.e., for $r \to 0$ because of
the short-range nature of the nuclear forces.

It should be mentioned that, strictly speaking, asymmetrical muonic
molecules ($Z_1 \neq Z_2$) do not form bound states but correspond to
resonant states of the continuous spectrum.
In this case an analogue of Eq.~(\ref{eq3}) is given in
Ref.~\cite{penko97} as
\begin{equation}
\label{eq4}
\lambda_{nf} = \frac{S}{(\pi M Z_1 Z_2)} \times \frac{1}{2l+1} \frac{M
 k_0}{4\pi} \Gamma \left| \Psi_{sc}(0) \right|^2 ,
\end{equation}
where $l$ is the orbital quantum number of the resonant state, $k_0$
is the relative momentum corresponding to the resonant energy,
$\Gamma$ is the width of the molecular state and $\Psi_{sc}(0)$ is the
wave function for the state of scattering at resonant energy.
In the limit of a very narrow resonance when $\Gamma \to 0$
Eqs.~(\ref{eq3}) and~(\ref{eq4}) coincide.
However, one should take into account the asymptotic part of the wave
function responsible for an in--flight fusion, including the possible
interferences between the resonant and nonresonant channels.

Let us briefly discuss the calculated nuclear fusion rates in the
$d\mu {}^{3}\mathrm{He}$ reaction presented in Table~\ref{tab:rates}.
The value given in Refs.~\cite{nagam89,kinox93} were given with some
references to a calculation by Kamimura but without any references to
the calculation method.
In Ref.~\cite{penko97} the author used a small variation basis and the
experimental value of the astrophysical factor $S \approx 6.32 \,
\mbox{MeV} \times \mbox{b}$ and found the nuclear fusion rate in the
$d\mu {}^{3}\mathrm{He}$ molecule in the $J = 0$ state to be $3.8
\times 10^6\ \mathrm{s}^{-1}$.

In Refs.~\cite{czapl96,czapl98} the nuclear fusion rate in the $d\mu
{}^{3}\mathrm{He}$ complex from the $J = 0$ state was calculated by
various methods.
Since the nuclear fusion rate in the $1s\sigma$ states of the $d\mu
{}^{3}\mathrm{He}$ molecule is much higher than the fusion rate form
the $2p\sigma$ state (because of a far smaller potential barrier), the
under-barrier $2p\sigma \rightarrow 1s\sigma$ transition was
calculated with finding the transition point in the complex $r$-plane.
This procedure is not quite unambiguous and therefore the nuclear
fusion rate in the $d\mu {}^{3}\mathrm{He}$ molecule was calculated in
an alternative way by reducing it to the $S$--factor and using
experimental data on low-energy scattering in ${}^{3}\mathrm{He} \,
(dp) \, {}^{4}\mathrm{He}$ reactions from Ref.~\cite{czapl96}.
However, the procedure of an approximation of the experimental data
for the ultralow energy region leads to some ambiguity of the results.
The results of the calculation by the above two methods may differ by a
factor of five for the $t\mu {}^3 \mathrm{He}$ molecule and by a
factor of three for the $d\mu {}^{3}\mathrm{He}$ molecule in
question~\cite{czapl98}.

The highest nuclear fusion rate was obtained in Ref.~\cite{harle89}.
Unlike the case in Ref.~\cite{czapl98}, where the barrier penetration
factor in the $2p\sigma \to 1s\sigma$ transition was evaluated, in
Ref.~\cite{harle89} the contribution from the $1s\sigma$ state to the
total wave function for the at small internuclear distances $r$ was
determined.
The determination of the contribution from this state to the total
mesomolecule wave function at small distance requires the solution of
a multichannel system of differential equations, which is a
complicated problem because of the singularity of the expansion
coefficients at small distances $r \to 0$.
As to the results of Ref.~\cite{bogda99} given in the last column of
Table~\ref{tab:rates}, it is difficult to judge the calculation method
used because the method for calculation the wave function at small
distances was not presented in the paper.

Different results of calculations of the fusion rate in the $d\mu
{}^{3}\mathrm{He}$ molecule reflect different approximations of the
solution to the Schr\"odinger equation for three particles with
Coulomb interaction.
The main uncertainty is associated with the results at small distances
and hence follows the spread of the calculated values for the nuclear
fusion rate in the $d\mu {}^{3}\mathrm{He}$ molecule given in
Table~\ref{tab:rates}.
When the adiabatic expansion is used, the important problem of
convergence of this expansion at small distances is usually ignored.
Such problems vanish if the direct solution of the Faddeev equations
in the configuration space is performed in
Refs.~\cite{kostr89,huxxx92,huxxx92b}.
For this reason the calculation of the fusion rate in the $d\mu
{}^{3}\mathrm{He}$ molecule using Faddeev equations in order to
adjudge discrepancies between different theoretical results becomes
very actual problem.

Much less has been done to study the nuclear fusion reaction in
the $d\mu {}^{3}\mathrm{He}$ experimentally.
The estimations of the lower limit for the fusion
reaction~(\ref{eq2}a) rate, has been done by a Gatchina -- PSI
collaboration using an ionization
chamber~\cite{balin92b,balin98,maevx99}.
Their results (see Table~\ref{tab:rates}) differ by several
orders of magnitude.
Another experiment aimed to measure the effective rate,
$\tilde{\lambda}_{f,p}$, of nuclear fusion reaction~(\ref{eq2}a) was
performed by our team~\cite{delro99}.
A preliminary result, also as estimation of lower limit, is shown in
Table~\ref{tab:rates}.

The purpose of this work was to measure the effective rate,
$\tilde{\lambda}_{f}$, of nuclear fusion reaction (\ref{eq2}a) in the
$d\mu {}^{3}\mathrm{He}$ complex with the formation of a 14.64~MeV
proton at two $\mathrm{D}_2 +{}^{3}\mathrm{He}$ mixture density
values.

\section{Measurement method}
\label{sec:method}

Figure~\ref{fig:scheme} shows a slightly simplified version of the
kinetics to be considered, when negative muons stop in the
$\mathrm{D}_2 +{}^{3}\mathrm{He}$ mixture.
The information on the fusion reaction~(\ref{eq2}a) rate in the $d \mu
{}^3 \mathrm{He}$ complex can be gained by measuring the time
distribution, $dN_p/dt$, and the total yield, $N_p$, of 14.64~MeV
protons.
These quantities are derived from the differential equations governing
the evolution of the $J=1,0$ states of the $d\mu {}^3\mathrm{He}$
molecules

Establishing the time dependence of the number of $d\mu
{}^{3}\mathrm{He}$ molecules, $N_{d\mu {}^3\mathrm{He}}^{J}(t)$, for
the two possible states $J$ is sufficient to predict the time spectrum
of the fusion products.
In the following, we will include the effective transition rate
$\tilde \lambda_{10}$ of the $d\mu {}^3\mathrm{He}$ complex between
the states $J=1$ and $J=0$.
The $\tilde \lambda_{10}$ transition is important if the $\tilde
\lambda_f^{1}$ and $\tilde \lambda_f^{0}$ rates differ strongly from
one to another, and an appropriate value of $\tilde \lambda_{10}$
permits the two rates to be measured.
This possibility can be checked by measuring the fusion using
different concentrations and densities which should also help clear up
the questions surrounding the mechanism of the~$\tilde \lambda_{10}$
transition~\cite{bystr99}, which is predicted to scale nonlinearly
with the density.

There is a direct transfer rate from ground state $d \mu$'s to
${}^{3}\mathrm{He}$'s but that rate is about 200 times smaller than
the $\lambda_{d {}^3\mathrm{He}}$ rate and will be
ignored~\cite{matve73}.
No hyperfine dependence on the $\lambda_{d {}^3\mathrm{He}}$ formation
rate is expected since the molecular formation involves an Auger
electron and bound state energies of many tens of electron
volts~\cite{arist81}.
Using the expectation that the $d \mu {}^3\mathrm{He}$ is formed
almost exclusively in the $J=1$ state, the solution for the fusion
products from the $J=0$ and $J=1$ states is relatively straightforward
given the $d \mu$ population.
The recycling of the muon after $d \mu {}^3\mathrm{He}$ fusion will be
ignored due to the extremely small probability of the fusion itself,
and thus the system of equations decouples into the $d \mu
{}^3\mathrm{He}$ sector, and the $dd$--fusion sector (where cycling
will be considered).
Since there is no expectation of a $J=0$ to $J=1$ transition, i.e.,
$\lambda_{01}$, the $d \mu {}^3\mathrm{He}${} sector is easily solved.

Formation of $d \mu d$ molecules from a $d \mu$ in hyperfine state
$F=3/2$ and $F=1/2$ is given by the effective rate $\tilde \lambda_F$,
whereas the branching ratio $\beta_F$ and sticking probability
$\omega_d$ model the number of muons lost from the cycle by sticking.
In both the initial condition on the number of $d \mu$ atoms, and in
the cycling efficiency after $dd$ fusion, $q_{1s}$ represents the
probability for a $d \mu$ atom formed in an excited state to reach the
ground state~\cite{bystr90e}.
Finally, $W_d$, represents the probability that the muon will be
captured by a deuterium atom given that there are both $\mathrm{D}_2$
and ${}^3\mathrm{He}$ in the mixture:
\begin{equation}
\label{eq4a}
      W_d = \frac{\mathrm{c}_d}{\mathrm{c}_d + A
      \mathrm{c}_{{}^3\mathrm{He}}} =
      \frac{X_{\mathrm{D}_2}}{X_{\mathrm{D}_2} + A'
      X_{{}^3\mathrm{He}}}
\end{equation}
where $\mathrm{c}_d$ and $\mathrm{c}_{{}^3\mathrm{He}}$ are the
deuterium and helium atomic concentration.
$A$ is the relative muon atomic capture probability by
a~${}^3\mathrm{He}$ atom compared to deuterium atom, and $A'$ is the
same ratio measured with respect to gas fraction concentrations ($X$).
An previous experimental measure exists for
$\mathrm{D}_2+{}^3\mathrm{He}$
$(A=1.7\pm0.2)$~\cite{bystr93,balin92b,bystr93c}, and theoretical
calculations for $A'$ have been made by J.~S.~Cohen~\cite{cohen99}:
for $\mathrm{D}_2+{}^3\mathrm{He}:\:A'=0.78$ and for
$\mathrm{HD}+{}^3\mathrm{He}:\:A'=0.68$.
Our gas mixtures have $\mathrm{c}_{{}^3\mathrm{He}} = 0.0496(10)$ and
thus $X_{{}^3\mathrm{He}}=0.0946(20)$.
By atomic concentration, and using the experimental value, we get
$W_d = 0.92(2)$.
Using theory and the gas fraction the result is the same, $W_d =
0.92$.
Using our own experiment~\cite{bystr94b}, $A=1.67^{+0.35}_{-0.33}$, to
determine $W_d$ leads also to the exact same value.

The differential equations governing the evolution of the $J=1,0$ spin
states of the $d \mu {}^3\mathrm{He}${} molecules are (see
~Fig.~\ref{fig:scheme}):
\begin{eqnarray}
      \frac{d\, N_{d \mu {}^3\mathrm{He}}^{1}}{dt}&=&+\varphi
      \mathrm{c}_{{}^3\mathrm{He}} \lambda_{d {}^3\mathrm{He}} N_{d
      \mu} -\lambda_{\Sigma}^1   N_{d \mu {}^3\mathrm{He}}^{1}
\label{equ:NdtHeO} \\
      \frac{d\, N_{d \mu {}^3\mathrm{He}}^{0}}{dt}&=&+\tilde
      \lambda_{10} N_{d \mu {}^3\mathrm{He}}^{1} -\lambda_{\Sigma}^0
      N_{d \mu {}^3\mathrm{He}}^{0}
\label{equ:NdtHeZ}
\end{eqnarray}
where $N_{d\mu}$ is the number of $d \mu$ atoms and with the
definition
\begin{eqnarray}
      \lambda_{\Sigma}^1   & = & \left(\lambda_0 + \lambda_p^{J=1} +
      \lambda_\gamma^{J=1} + \lambda_e^{J=1} + \lambda_f^{J=1}
      \right)
\label{equ:lamsumdefO} \\
      \lambda_{\Sigma}^0   & = & \left(\lambda_0 + \lambda_p^{J=0} +
      \lambda_\gamma^{J=0} + \lambda_e^{J=0} + \lambda_f^{J=0} \right)
      \, ,
\label{equ:lamsumdefZ}
\end{eqnarray}
and
\begin{eqnarray}
  \label{eq:ldmu}
  \lambda_{d \mu} & = & \lambda_0 + \varphi
  \mathrm{c}_{{}^3\mathrm{He}} \lambda_{d {}^3\mathrm{He}} \nonumber
  \\
  & + & \varphi \mathrm{c}_d \tilde \lambda_F \left[1 - W_d q_{1s} (1
  - \beta_F \omega_d) \right ] \, .
\end{eqnarray}
The yield for protons between two given times after the muon arrival,
$t_1$ and $t_2$, is:
\begin{eqnarray}
       Y_p(t_1,t_2) & = & Y_p^1(t_1,t_2) +Y_p^0(t_1,t_2) \nonumber \\
       & = & N_\mu^\mathrm{D/He} \cdot \frac{\tilde \lambda_f}
       {\lambda_\Sigma} \frac{\varphi \mathrm{c}_{{}^3\mathrm{He}}
       \lambda_{d {}^3\mathrm{He}} W_d q_{1s}
       \varepsilon_Y \varepsilon_p}{\lambda_{d\mu}} \: ,
\label{equ:pyieldt}
\end{eqnarray}
where the difference in time exponents has been defining as the yield
efficiency:
\begin{equation}
      \varepsilon_Y = \left(e^{\lambda_{d\mu} t_1} - e^{\lambda_{d\mu}
      t_2} \right) \: .
\label{equ:effY}
\end{equation}
and with the effective fusion rate defined as
\begin{eqnarray}
      \tilde \lambda_f & = & \left(
      {\lambda_f^{J=1}}\frac{\lambda_{\Sigma}^0 }{\tilde \lambda_{10}
      +\lambda_{\Sigma}^0 } + \lambda_f^{J=0} \frac{\tilde
      \lambda_{10}}{\tilde \lambda_{10}+\lambda_{\Sigma}^0 } \right)
\label{equ:efffus} \\
      \lambda_\Sigma & = & \lambda_{\Sigma}^0 \left(\frac{\tilde
      \lambda_{10} + \lambda_{\Sigma}^1  }{\tilde \lambda_{10} +
      \lambda_{\Sigma}^0 }\right).
\label{equ:effdec}
\end{eqnarray}
In the above equations, $N_\mu^\mathrm{D/He}$ is the number of muons
stopped in the $\mathrm{D}_2 +{}^{3}\mathrm{He}$ mixture and $\varphi$
is the mixture atomic density relative to the liquid hydrogen density
(LHD, $N_0 = 4.25 \times 10^{22} \, \mbox{cm}^3$).

When protons are detected in coincidences with muon decay electrons,
later on called the del-$e$ criterion, the fusion rate from
Eq.~(\ref{equ:pyieldt}) takes the form
\begin{equation}
\label{eq15}
      \tilde \lambda_f = \frac{ Y_p(t_1,t_2) \lambda_{d\mu}
      \lambda_\Sigma } {N_\mu^\mathrm{D/He} W_d\, q_{1s} \, \varphi \,
      \mathrm{c}_{{}^3\mathrm{He}} \lambda_{d {}^3\mathrm{He}}
      \varepsilon_p \, \varepsilon_e \, \varepsilon_t \, \varepsilon_Y
      }\, ,
\end{equation}
where $\varepsilon_e$ is the detection efficiency for muon decay
electrons and $\varepsilon_t$ defined as
\begin{equation}
  \label{epsilont}
       \varepsilon_t = e^{-\lambda_0 t_{ini}} - e^{-\lambda_0
       t_{fin}}
\end{equation}
is the time efficiency depending on the interval during which we
accept the muon decay electrons.
Note that Eqs.~(\ref{equ:pyieldt}--\ref{eq15}) are valid when the
proton detection times are $t \gg 1/ \lambda_\Sigma$.
The values $\varepsilon_p$ and $\lambda_\Sigma$ are found through
calculation.
Note an important feature of this experimental setup: $\tilde
\lambda_f$\ is found by using the experimental values of
$\lambda_{d\mu}$, $\varepsilon_e$, $W_d$ , $\lambda_{d
{}^3\mathrm{He}}$, and $q_{1s}$.

The information on these quantities corresponds to the conditions of a
particular experiment and is extracted by the analysis of yields and
time distributions of the 6.85~keV $\gamma$~rays from
reaction~(\ref{eq1}), prompt and delayed x~rays of $\mu
{}^{3}\mathrm{He}$ atoms in the $\mathrm{D}_2 +{}^{3}\mathrm{He}$
mixture and muon decay electrons.
The quantity $\lambda_{d {}^3\mathrm{He}}$ is determined from
Eq.~(\ref{eq:ldmu}) where $\beta_F= 0.58$, $\omega_d= 0.122(3)$ are
taken from Refs.~\cite{balin84}.
$\tilde \lambda_F = 0.05 \times 10^6 \, \mbox{s}^{-1}$ is taken from
Ref.~\cite{petit96}.
The rate $\lambda_{d\mu}$ is the slope of the time distribution of
$\gamma$~ray from reaction (\ref{eq1}).

The procedure of measuring $q_{1s}$, $\lambda_{d {}^3\mathrm{He}}$,
$W_d$ , $\varepsilon_e$, $A$ and $\lambda_\gamma$ (the partial
probability for the radiative $d\mu {}^3\mathrm{He}$ complex decay
channel) as well as our results are described in detail in our
previous work~\cite{bystr04}.

\section{Experimental setup}
\label{sec:setup}

The experimental layout (see Fig.~\ref{fig:apparatus}) was described
in details in Refs.~\cite{borei98,bystr04,bystr04b}.
The experimental facility was located at the $\mu$E4 beam line of the
PSI meson factory (Switzerland) with the muon beam intensity around $2
\times 10^4 \, \mbox{s} ^{-1}$.
After passing through a thin plastic entrance monitoring counter muons
hit the target and stopped there initiating a sequence of processes
shown in Fig.~\ref{fig:scheme}.
The electronics are protected from muon pileup within a $\pm 10 \,
\mu$s time gate so pileup causes a 30\% reduction in the effective
muon beam.
Thus, we have a number of ``good muons'', called $N_\mu$, stopping in
our target.

\begin{figure}[t]
  \centerline{
\setlength{\unitlength}{0.1mm}
\begin{picture}(800,800)(0,0)
\put(390,450){\oval(200,200)}
\put(380,440){T}
\put(250,300){\line(1,0){300}}
\put(250,600){\line(1,0){300}}
\put(550,600){\line(0,-1){300}}
\put(250,600){\line(0,-1){120}}
\put(250,300){\line(0,1){120}}
\put(250,420){\line(1,0){40}}
\put(250,480){\line(1,0){40}}
\put(290,480){\line(0,-1){60}}
\put(260,520){V}
\put(130,425){\framebox(150,50)}
\put(150,438){\small Ge$_S$}
\put(300,565){\framebox(200,15)}
\put(300,570){\line(1,0){200}}
\put(130,680){\small Si$_{UP}$}
\put(220,670){\vector(1,-1){90}}
\put(300,320){\framebox(200,15)}
\put(300,330){\line(1,0){200}}
\put(130,220){\small Si$_{DO}$}
\put(220,240){\vector(1,1){80}}
\put(515,350){\framebox(15,200)}
\put(520,350){\line(0,1){200}}
\put(630,630){\small Si$_{RI}$}
\put(620,620){\vector(-1,-1){90}}
\put(220,625){\framebox(350,15)}
\put(630,715){\small E$_{UP}$}
\put(620,705){\vector(-1,-1){60}}
\put(580,290){\framebox(15,320)}
\put(670,205){\small E$_{RI}$}
\put(660,225){\vector(-1,1){60}}
\put(220,260){\framebox(350,15)}
\put(600,150){\small E$_{DO}$}
\put(620,190){\vector(-1,1){60}}
\put(220,290){\framebox(15,120)}
\put(220,490){\framebox(15,120)}
\put(100,565){\small E$_{LE}$}
\put(180,550){\vector(1,-1){40}}
\put(250,660){\line(1,0){300}}
\put(250,660){\line(0,1){100}}
\put(550,660){\line(0,1){100}}
\put(355,700){\small NE213}
\put(610,300){\line(1,0){100}}
\put(610,600){\line(1,0){100}}
\put(610,600){\line(0,-1){300}}
\put(630,435){\small Ge$_M$}
\put(250,240){\line(1,0){300}}
\put(250,240){\line(0,-1){100}}
\put(550,240){\line(0,-1){100}}
\put(380,180){\small Ge$_B$}
\put(250,80){\line(1,0){300}}
\put(250,75){\line(0,1){10}}
\put(240,30){0}
\put(400,75){\line(0,1){10}}
\put(550,75){\line(0,1){10}}
\put(520,30){10 cm}
\end{picture}
}
  \caption{Apparatus used in the $\mu$E4 area.  The view is that of
    the incoming muon.  Note that the T1 and T0 scintillators are not
    shown.  The labels are explained in the text.}
  \label{fig:apparatus}
\end{figure}

Three pairs of Si($dE-E$) telescopes were installed directly behind
135~$\mu$m thick kapton windows and a 0.17~cm$^3$ germanium detector
behind a 55~$\mu$m thick kapton window to detect the 14.64~MeV protons
from reaction~(\ref{eq2}a) and the 6.85~keV $\gamma$~rays from
reaction~(\ref{eq1}), respectively.
The Si telescopes with a 42~mm diameter were made of a 4~mm thick
Si($E$) detector and a thin, 360~$\mu$m thick, Si($dE$) detector,
respectively.
An assembly of Si detectors like that give a good identification of
protons, deuterons, and electrons based on different energy losses of
the above particles in those detectors.
Muon decay electrons were detected by four pairs of scintillators,
E$_{UP}$, E$_{DO}$, E$_{RI}$ and E$_{LE}$, placed around the vacuum
housing of the target.
The total solid angle of the electron detectors was $\approx 17$\%.
The cryogenic target was located inside the vacuum housing.
The design of the target is described in detail in
Refs.~\cite{stolu99,borei98}.

The analysis of the 6.85~keV $\gamma$--ray time distributions allows
us to determine the disappearance rate, $\lambda_{d\mu}$, for the
$d\mu$ atoms in the $\mathrm{D}_2 +{}^{3}\mathrm{He}$ mixture.
Note that the presence of a signal from the electron detectors during
a certain time interval (the del-$e$ criterion) whose beginning
corresponds to the instant of time when the $K\alpha$, $K\beta$, and
$K\gamma$ lines of $\mu \mathrm{He}$ atoms is detected makes it
possible to determine uniquely the detection efficiency for muon decay
electrons.
When the del-$e$ criterion is used in the analysis of events detected
by the Si($dE-E$) telescopes one obtains a suppression factor of
$300-400$ of the background, which is quite enough to meet the
requirements of the experiment on the study of nuclear fusion in the
$d\mu {}^3\mathrm{He}$  complex.

Our experiment included two runs with the $\mathrm{D}_2
+{}^{3}\mathrm{He}$ mixture.
The experimental conditions are listed in Table~\ref{tab:conditions}.
In addition, we performed different measurements with pure
$\mathrm{D}_2$, ${}^3\mathrm{He}$, and ${}^4\mathrm{He}$ at different
pressures and temperature.
Details are given in Refs.~\cite{bystr04b}.

\begin{table}[ht]
\begin{ruledtabular}
      \caption{Experimental conditions for the $\mathrm{D}_2
      +{}^{3}\mathrm{He}$ mixtures with an atomic concentration of
      helium $\mathrm{c}_{{}^3\mathrm{He}} = 0.0496$. $N_\mu$ is the
      number of muon stopped in our apparatus.}
\label{tab:conditions}
\begin{tabular}{cccccc}
Run & P$_\mu$& T & p & $\varphi$ & $N_{\mu}$ \\
& [MeV/c]& [K] & [kPa] &[LHD]& [ $10^9$ ] \\ \hline
I & 34.5 & 32.8 & 513.0 & 0.0585 & 8.875 \\
II & 38.0 & 34.5 & 1224.4 & 0.1680 & 3.928 \\
\end{tabular}
\end{ruledtabular}
\end{table}

The germanium detector was calibrated using $^{55}$Fe and $^{57}$Co
sources.
The Si($dE-E$) detectors were calibrated using a radioactive
$^{222}$Rn source.
Before the cryogenic target was assembled, a surface saturation of the
Si($dE$) and Si($E$) detectors by radon was carried out.
The $^{222}$Rn decay with the emission of alpha-particles of energies
5.3, 5.5, 6.0, and 7.7~MeV were directly detected by each of the Si
detectors.
The linearity of the spectrometric channels of the Si detectors in the
region of detection of protons with energies $8-15$~MeV was checked
using exact-amplitude pulse generators.

\section{Analysis of the experimental data}
\label{sec:analysis}

\subsection{Determination of the $d\mu {}^{3}\mathrm{He}$complex formation rate}
\label{sec:formrate}

\begin{figure}[ht]
\includegraphics[angle=90,width=\linewidth]{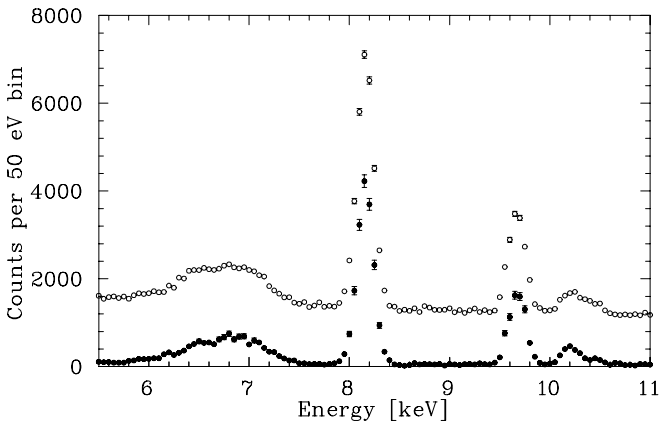}
                \caption{Energy spectra of the events detected by the
                Ge$_S$ detector in run I without (open circles) and
                with (full circles) the del-$e$ criterion.}
\label{fig:energy-ge}
\end{figure}

By way of example Fig.~\ref{fig:energy-ge} shows energy spectra of
events detected by the germanium detector in run I without and with
the del-$e$ criterion.
The rather wide left peak corresponds to the $\gamma$~rays with an
average energy of 6.85~keV and the three right peaks correspond to the
$K\alpha$, $K\beta$, $K\gamma$ lines of $\mu \mathrm{He}$ atoms with
energies 8.17, 9.68, and 10.2~keV, respectively.
As seen in Fig. \ref{fig:energy-ge}, the suppression factor for the
background detected by the germanium detector with the del-$e$
criterion is of the order of $10^3$.

\begin{figure}[hb]
\includegraphics[angle=90,width=\linewidth]{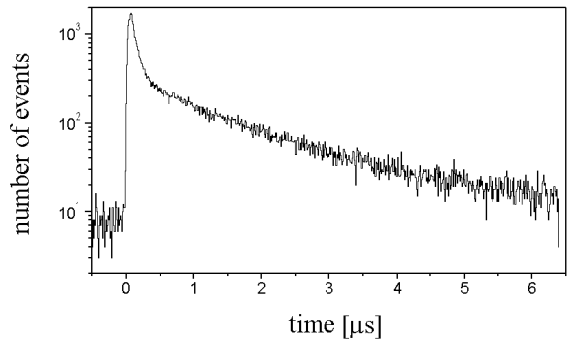}
                \caption{Time distributions of the 6.85~keV
                $\gamma$--quanta resulting from radiative
                de-excitation of the $d\mu {}^ {3}\mathrm{He}$ complex
                obtained in coincidences with the muon decay electrons
                in run I\@. }
\label{fig:time-ge}
\end{figure}

Figure~\ref{fig:time-ge} shows time distributions of 6.85~keV
$\gamma$~rays resulting from radiative de-excitation of the $d\mu
{}^3\mathrm{He}$ complex in runs I and II\@.
The distributions were measured in coincidences with delayed muon
decay electrons.
The experimental time distributions of $\gamma$~rays shown in
Fig.~\ref{fig:time-ge} were approximated by the following expression
\begin{equation}
\label{eq18}
      \frac{dN_\gamma}{t} = B^\gamma e^{-\lambda_{d \mu} t} +
      C^\gamma e^{- \lambda_0 t} + D^\gamma \, ,
\end{equation}
where $B^\gamma$, $C^\gamma$, and $D^\gamma$ are the normalization
constants.
The second and third terms in Eq.~(\ref{eq18}) describe the
contribution from the background.
The analysis of the time distributions of the 6.85~keV $\gamma$~rays
yielded values of $\lambda_{d\mu}$ and thus the formation rates
$\lambda_{d {}^3\mathrm{He}}$.
Results are given in Table~\ref{tab:param}.

\begin{table}[b]
\begin{ruledtabular}
      \caption{Parameters used to determine the formation rates
      $\lambda_{d {}^3\mathrm{He}}$. The value $W_d = 0.92(2)$ was
      used for both runs.}
\label{tab:param}
\begin{tabular}{cccc}
Run & $q_{1s}$ & $\lambda_{d \mu}$ & $\lambda_{d {}^3\mathrm{He}}$ \\
& & [$10^6$ s$^{-1}$] & [$10^6$ s$^{-1}$] \\ \hline
I & 0.882 (18) & $1.152 (36)_{stat} (30)_{syst}$ & $240
(13)_{stat}(15)_{syst}$ \\
II& 0.844 (20) & $2.496 (58)_{stat} (100)_{syst}$ & $244
(6)_{stat}(16)_{syst}$ \\
\end{tabular}
\end{ruledtabular}
\end{table}

The systematic error is larger than the uncertainty of the result
caused by various possible model of the background, including the case
where it is equal to zero (e.g., when time structure of the background
is inaccurately known).
We describe the procedure of determining $\lambda_{d {}^3\mathrm{He}}$
in more detail in Ref.~\cite{czapl96}.

\subsection{Number of muon stops in the $\mathrm{D}_2 +{}^{3}\mathrm{He}$ mixture}
\label{sec:mstops}

The number of muon stops in the $\mathrm{D}_2 +{}^{3}\mathrm{He}$
mixture was determined by analyzing time distributions of events
detected by the four electron counters,
We detailed this matter in Refs.~\cite{czapl96c,borei98,bystr04b}.
Here it is pertinent to dwell upon some particular points in
determination of this value.

\begin{figure}[ht]
\includegraphics[width=\linewidth]{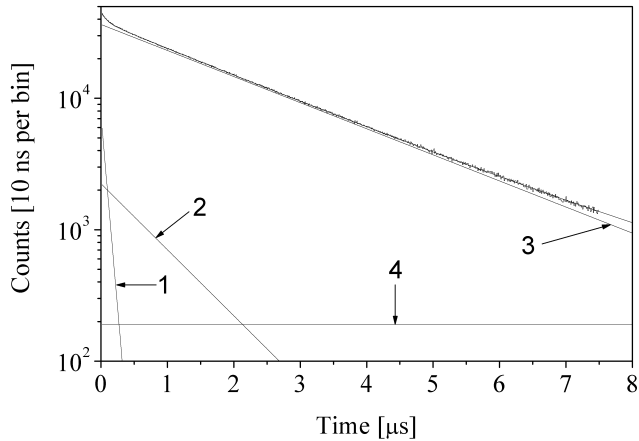}
                \caption{Time distributions of muon decay electrons
                measures in run I\@. The solid curves are the results
                of fitting its components (see Eq.~(\ref{eq19a})): 1 -
                Au; 2 - Al; 3 - $\mathrm{D}_2 +{}^{3}\mathrm{He}$; 4 -
                constant background.  }
\label{fig:time-dele}
\end{figure}

By way of example Fig.~\ref{fig:time-dele} shows the time distribution
of muon decay electrons measured in run I\@.
To determine the number of muon stops in the $d\mu {}^3\mathrm{He}$
the time distribution of the detected electrons, $dN_e/dt$, is
approximated by an expression which is superposition of four exponents
and a background of accidental coincidences
\begin{equation}
      \label{eq19}
      \frac{dN_e }{dt} = A_\mathrm{Al}^e e^ { - \lambda _{\mathrm
      {Al}} t} + A_\mathrm{Au}^e e^ { - \lambda _{\mathrm {Au}} t} +
      A_\mathrm{He}^e e^ { - \lambda _{\mathrm {He}} t} +
      A_\mathrm{D}^e e^ { - \lambda _0 t} +
      B^e,
\end{equation}
where $A_\mathrm {Al}^e$, $A_\mathrm {Au}^e$, $A_\mathrm {He}^e$ and
$A_\mathrm {D}^e$, are the normalized amplitudes with
\begin{equation}
      \label{eq20}
       A^e_i = N_\mu^i Q_i \lambda_0 \varepsilon_e \qquad i=
       \mathrm{Al},  \, \mathrm{Au}, \, \mathrm{He}, \, \mathrm{D},
\end{equation}

and
\begin{eqnarray}
      \label{eq20a}
      \lambda _{\mathrm {Al}} & = & Q_{\mathrm {Al}} \cdot \lambda_0 +
      \lambda_\mathrm{cap}^{\mathrm {Al} }\, , \nonumber \\
      \lambda _{\mathrm {Au}} & = & Q_{\mathrm {Au}} \cdot \lambda_0 +
      \lambda_\mathrm{cap}^{\mathrm {Au} }\, , \\
      \lambda _{\mathrm {He}} & = & \lambda_0 +
      \lambda_\mathrm{cap}^{\mathrm {He} }\, , \nonumber \\
\end{eqnarray}
are the muon disappearance rates in the different elements (the rates
are the inverse of the muon lifetimes in the target wall materials).
In reality, Eq.~(\ref{eq19}) is an approximation of a more complex
equation, which can be found in Ref.~\cite{knowl99b}.
The different rates are $\lambda_0 = 0.455 \times 10^6 \,
\mbox{s}^{-1}$ and $\lambda_{cap}^{\mathrm{He}} = 2216(70) \,
\mbox{s}^{-1}$~\cite{maevx96}.
The nuclear capture rates in aluminum and gold,
$\lambda_\mathrm{cap}^{\mathrm {Al}}= 0.7054 (13) \times 10^6 \,
\mbox{s}^{-1}$ and $\lambda_\mathrm{cap}^{\mathrm {Au}}= 13.07 (28)
\times 10^6 \, \mbox{s}^{-1}$, are taken from Ref.~\cite{suzuk87}.
$Q_{\mathrm{Al}}$ and $Q_{\mathrm{Au}}$ are the Huff factors, which
take into account that muons are bound in the $1s$ state of the
respective nuclei when they decay.
This factor is negligible for helium but necessary for aluminum
$Q_{\mathrm {Al}}=0.993$ and important for gold
$Q_{\mathrm{Au}}=0.850$~\cite{suzuk87}.
The constant $B^e$ characterizes the random coincidence background.

We denote $N_\mu$ as the total number of muon stops in
the target, $N_\mu^\mathrm{Al}$, $N_\mu^\mathrm{Au}$, and
$N_\mu^\mathrm{D/He}$ as the numbers of muon stops in Al, Au, and the
gaseous $\mathrm{D}_2 +{}^{3}\mathrm{He}$ mixture, respectively.
Thus, we have the relation
\begin{equation}
  \label{eq25a}
      N_\mu = N_\mu^\mathrm{Al} + N_\mu^\mathrm{Au} +
      N_\mu^\mathrm{D/He} \, .
\end{equation}

Since the muon decay with emission electrons in the $\mathrm{D}_2
+{}^{3}\mathrm{He}$ mixture take place from the $1s$ state of the
$d\mu$ or $\mu{}^3 \mathrm{He}$ atom, the third and fourth terms in
Eq(\ref{eq19}) will differ only by the values of the amplitudes
$A^e_\mathrm{He}$ and $A^e_\mathrm{D}$ because the slopes of both
exponents are practically identical ($\lambda_\mathrm{He} = 0.457 \,
\mu \mbox{s}^{-1}$, $\lambda_{0} = 0.455\ \mu \mathrm{s}^{-1}$).
In this connection the following simplified expression was used to
approximate experimental time distributions of
\begin{equation}
      \label{eq19a}
      \frac{dN_e }{dt} = A_\mathrm{Al}^e e^ { - \lambda _{\mathrm
      {Al}} t} + A_\mathrm{Au}^e e^ { - \lambda _{\mathrm {Au}} t} +
      A_\mathrm{D/He}^e e^ { - \tilde \lambda _{\mathrm {D/He}} t} +
      B^e,
\end{equation}
Under our experimental conditions of runs I and II, we obtained the
effective rates $\tilde \lambda_{D/He} = 0.4563 \, \mu \mbox{s}^{-1}$
and 0.4567~$\mu \mbox{s}^{-1}$, respectively.
With these effective muon decay rates, the uncertainty in the
calculated number of muon stops in the gaseous $\mathrm{D}_2
+{}^{3}\mathrm{He}$ mixture is negligibly small as compared with the more
rigorous calculation of this value by Eq.~(\ref{eq19}).

The amplitudes in Eq.~(\ref{eq20}) are expressed in terms of the
factors $a_{Al}$, $a_{Au}$, and $a_{D/He}$, defined as the partial
muon stopping in Al, Au, and $\mathrm{D}_2 +{}^{3}\mathrm{He}$
mixture,
\begin{equation}
\label{eq26}
      a_{i} = \frac{N_\mu^i}{N_\mu} \, , \qquad \sum_i a_i = 1 \,
      \qquad i = \mathrm{Al}, \, \mathrm{Au}, \, \mathrm{D/He} \, ;
\end{equation}
take the new form
\begin{equation}
\label{eq27}
      A^e_{i} = N_\mu \lambda_{0} Q_i \varepsilon_e a_{i}\, .
\end{equation}

The electron detection efficiency, $\varepsilon_e$, of the detectors
E$_{UP}$, E$_{DO}$, E$_{RI}$ and E$_{LE}$ was determined as a ratio
between the number of events belonging to the $K$--lines of the $\mu
{}^{3}\mathrm{He}$ atoms, found from the analysis of the data with and
without the del-$e$ criterion,
\begin{equation}
\label{eq29}
      \varepsilon_e = \frac{N_{x-e}}{N_x}\, ,
\end{equation}
where $N_{x-e}$ and $N_x$ are the numbers of events belonging to
$K$--lines of the $\mu {}^{3}\mathrm{He}$ atom and detected by the
germanium detector with and without coincidence with the electron
detectors.
The thus measured experimental value is electron detection efficiency
averaged over the target volume.
Table~\ref{tab:effi} presents the results.

\begin{table}[ht]
\begin{ruledtabular}
      \caption{Electron detection efficiencies, $\varepsilon_e$, in
      [\%].}
\label{tab:effi}
\begin{tabular}{cccccc}
Run  & \multicolumn{5}{c}{Detector} \\
     & E$_{UP}$ & E$_{RI}$ & E$_{DO}$ & E$_{LE}$ & all \\\hline
I  & 4.77(16)  & 5.69(16)  & 4.91(16)  & 0.169(24) & 16.40(31) \\
II & 4.53(15)  & 5.89(18)  & 4.88(14)  & 0.114(39) & 16.34(39) \\ \hline
$\varepsilon_e$ & 4.65(12) & 5.79(12) & 4.89(12) & 0.148(23) &
16.37(22) \\
\end{tabular}
\end{ruledtabular}
\end{table}

The electron detection efficiency of the detector E$_{LE}$ is
considerably lower than that of each of the other three electron
detectors.
This is because the material (Al, Fe) layer which the muon decay
electron has to pass through in the direction of the detector E$_{LE}$
is thicker than material layers in the direction of the other electron
detectors.

Table~\ref{tab:fraction} lists the values of the fraction of muons
stopped in the $\mathrm{D}_2 +{}^{3}\mathrm{He}$ mixture, $a_{D/He}$,
found from the analysis of the time distributions of the events
detected by the four electron detectors in runs I and II\@.
Note that when the $a_{D/He}$ fraction, was calculated by
Eqs.~(\ref{eq26}) and~(\ref{eq27}) it was assumed that the electron
detection efficiency by each of the detectors E$_{UP}$, E$_{DO}$,
E$_{RI}$ and E$_{LE}$ did not depend on the coordinates of the muon
stop point in the target (be it in the target walls or in the
$\mathrm{D}_2 +{}^{3}\mathrm{He}$ mixture).

\begin{table}[ht]
\begin{ruledtabular}
      \caption{Fraction of muons stopped in the gaseous
      deuterium-helium mixture found by the absolute
      method. $N_\mu^{D/He}$ is the number of muons stops in the
      $\mathrm{D}_2 +{}^{3}\mathrm{He}$ gas mixture.}
\label{tab:fraction}
\begin{tabular}{ccc}
Run  & $a_{D/He}$ & $N_\mu^{D/He}$ \\
     & [\%]           & [$10^{9}$] \\ \hline
I  & $47.5 (6)_{stat} (30)_{syst} $ & 4.216 \\
II & $66.6 (10)_{stat} (39)_{syst}$ & 2.616 \\
\end{tabular}
\end{ruledtabular}
\end{table}

The systematic errors were determined as one half of the maximum
spread between the $a_{D/He}$ values found from analysis of the time
distributions of the electrons detected by each of the electron
detectors E$_{UP}$, E$_{DO}$, E$_{RI}$ and E$_{LE}$.
Note that the fraction of muons stopped in gas, $a_{D/He}$, is a
result of simultaneously fitting all time distributions obtained with
each of the electron detectors (and not a result of averaging all four
distributions corresponding to each of the four detectors).

\subsection{Determination of the detection efficiency for 14.64~MeV protons}
\label{sec:effi}

To determine the proton detection efficiency, $\varepsilon_p$, of the
three Si($dE-E$) telescopes, one should know the distribution of muon
stops over the target volume in runs I and II\@.
The average muon beam momentum $\overline{\mathrm{P}}_\mu$
corresponding to the maximum fraction $a_{D/He}$ of muons stopped in
the $\mathrm{D}_2 +{}^{3}\mathrm{He}$ mixture in runs I and II was
found by varying the muon beam momentum $\overline{\mathrm{P}}_\mu$
and analyzing the time distributions of the detected electrons by
Eq.~(\ref{eq19a}).
Next, knowing the average momentum $\overline{\mathrm{P}}_\mu$ and the
beam momentum spread, we simulated the real distribution of muon stops
in runs I and II by the Monte Carlo (MC) method~\cite{jacot97b}.
The results of the simulation were used in another Monte Carlo program
to calculate the detection efficiency of each pair of Si($dE-E$)
detectors for protons from reaction (\ref{eq2}a)~\cite{wozni96}.
The algorithm of the calculation program included simulation of the
muon stop points in the $\mathrm{D}_2 +{}^{3}\mathrm{He}$ mixture and
the $d\mu$ and $\mu {}^{3}\mathrm{He}$ atom formation points, the
consideration of the entire chain of processes occurring in the
$\mathrm{D}_2 +{}^{3}\mathrm{He}$ mixture from the instant when the
muon hits the target to the instant of possible production of
14.64~MeV protons in the fusion reaction in the $d\mu
{}^{3}\mathrm{He}$ complex.
The calculation program took into account the proton energy loss in
the gas target, kapton windows and Si($dE-E$) detectors themselves (in
the thin Si($dE$) and thick Si($E$) detectors).
The proton detection efficiency $\varepsilon_p$ was calculated at the
$q_{1s}$, $W_d$, and $\lambda_{d\mu}$ values (see
Table~\ref{tab:param}) corresponding to our experimental conditions.
The scattering cross sections of $d\mu$ atoms form $\mathrm{D}_2$
molecules were taken from Refs.~\cite{chicc92,adamc96,melez92}.

We ceased tracing the muon stopped in the target when
\begin{itemize}
      \item[a)] the muon decays ($\mu^- \to
      e^-+\nu_\mu+\tilde{\nu}_e$)
      \item[b)] the muon is transferred from the deuteron to the
      ${}^{3}\mathrm{He}$ nucleus with the formation of a
      ${}^{3}\mathrm{He}$$\mu$ atom
      \item[c)] nuclear fusion occurs in the $d\mu {}^{3}\mathrm{He}$
      complex
      \item[d)] a $dd\mu \to p + t + \mu$ reaction proceeds in the
      $dd\mu$ molecule.
\end{itemize}
Note that the algorithm of the program also involved the consideration
of the background process resulting from successive occurrence of the
reactions
\begin{eqnarray}
\label{eq31}
      d\mu+d\rightarrow dd\mu \rightarrow\! &\! ^3\mathrm{He}\!
      &\!(0.8\ \mathrm{MeV}) +n\nonumber\\
      & +\nonumber\\
      & d\! &\!  \to \alpha+ p (14.64 \, \mathrm{MeV}).
\end{eqnarray}
This reaction~(\ref{eq31}) is called $d {}^3 \mathrm{He}$ ``fusion in
flight''.

\begin{figure}[ht]
\includegraphics[angle=90,width=\linewidth]{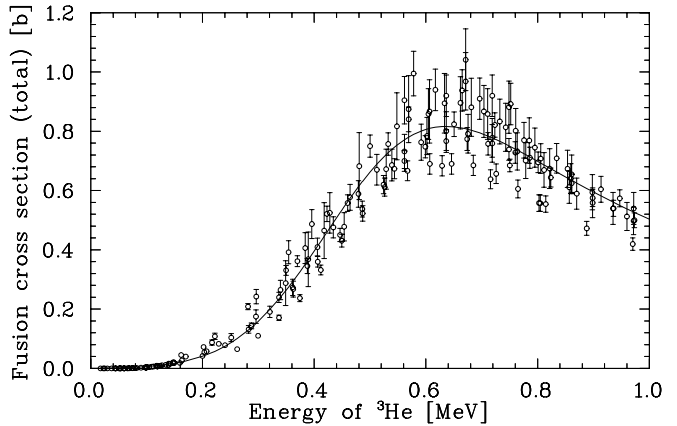}
                \caption{Cross section for the reaction
                ${}^{3}\mathrm{He} + d \to {}^{4}\mathrm{He} + p$ in
                flight (reaction~(\ref{eq31})) as a function of the
                ${}^{3}\mathrm{He}$ nucleus-deuteron collision
                energy. The solid curve is the result of averaging the
                entire bulk of the presented experimental data.  }
\label{fig:xs}
\end{figure}

In our calculations we used the dependence of the cross section for
reaction~(\ref{eq31}) on the ${}^{3}\mathrm{He}$ deuteron collision
energy, averaged over the data of
Refs.~\cite{white97,kunzx55,kljuc56,allre52,argox52,freie54}.
Figure~\ref{fig:xs} displays the cross section dependence on the
${}^{3}\mathrm{He}$ deuteron collision energy.
The program also took into account the energy loss of
${}^{3}\mathrm{He}$ nuclei in the $\mathrm{D}_2 +{}^{3}\mathrm{He}$
mixture caused by ionization of ${}^{3}\mathrm{He}$ atoms and
deuterium molecules.
The time distributions of protons from reactions (\ref{eq2}a)
and~(\ref{eq31}) under the same experimental conditions have
completely different shapes in accordance with the kinetics of
processes in the $\mathrm{D}_2 +{}^{3}\mathrm{He}$ mixture.

\begin{figure*}[thb]
\includegraphics[width=0.35\linewidth,angle=90]{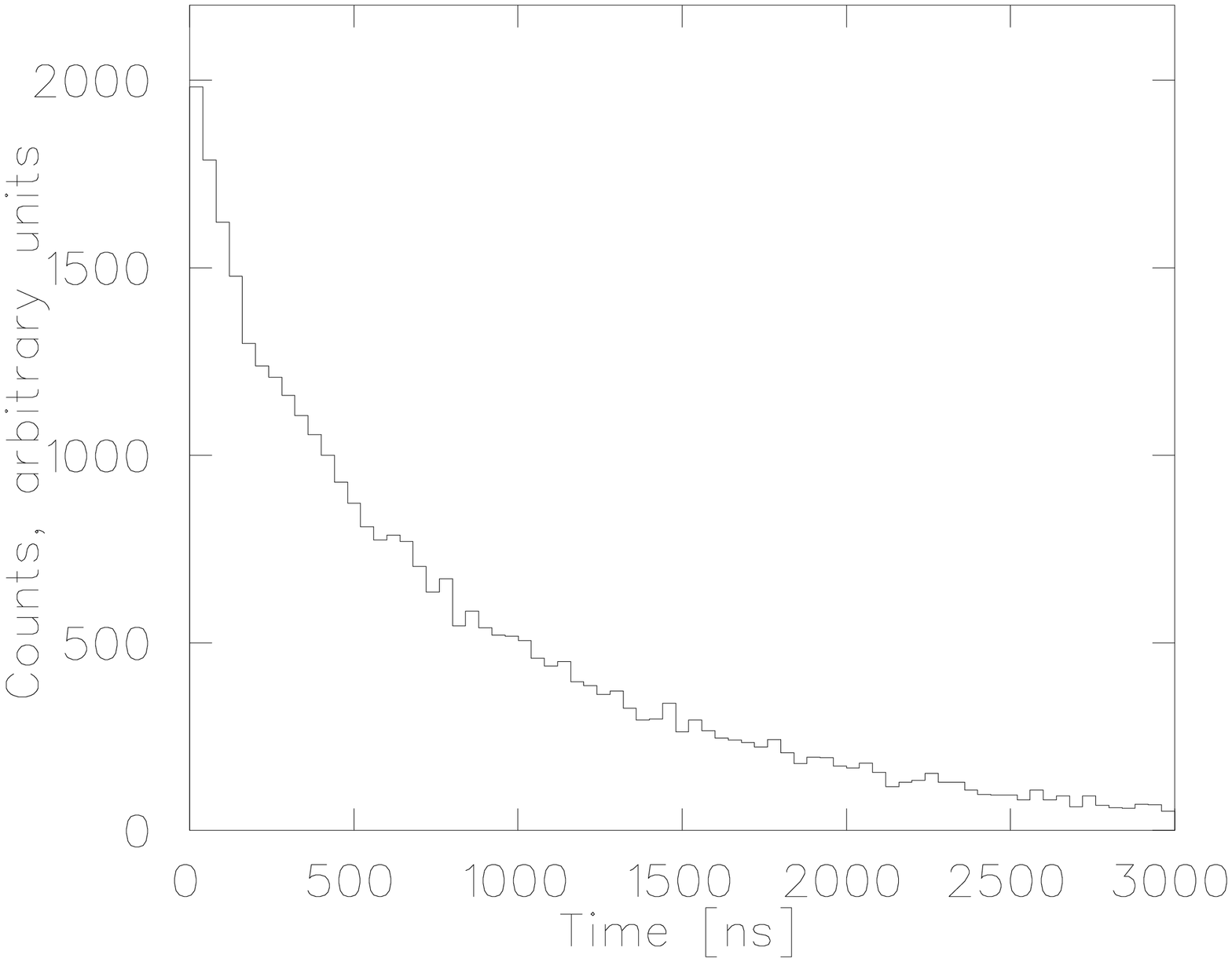}
\includegraphics[width=0.35\linewidth,angle=90]{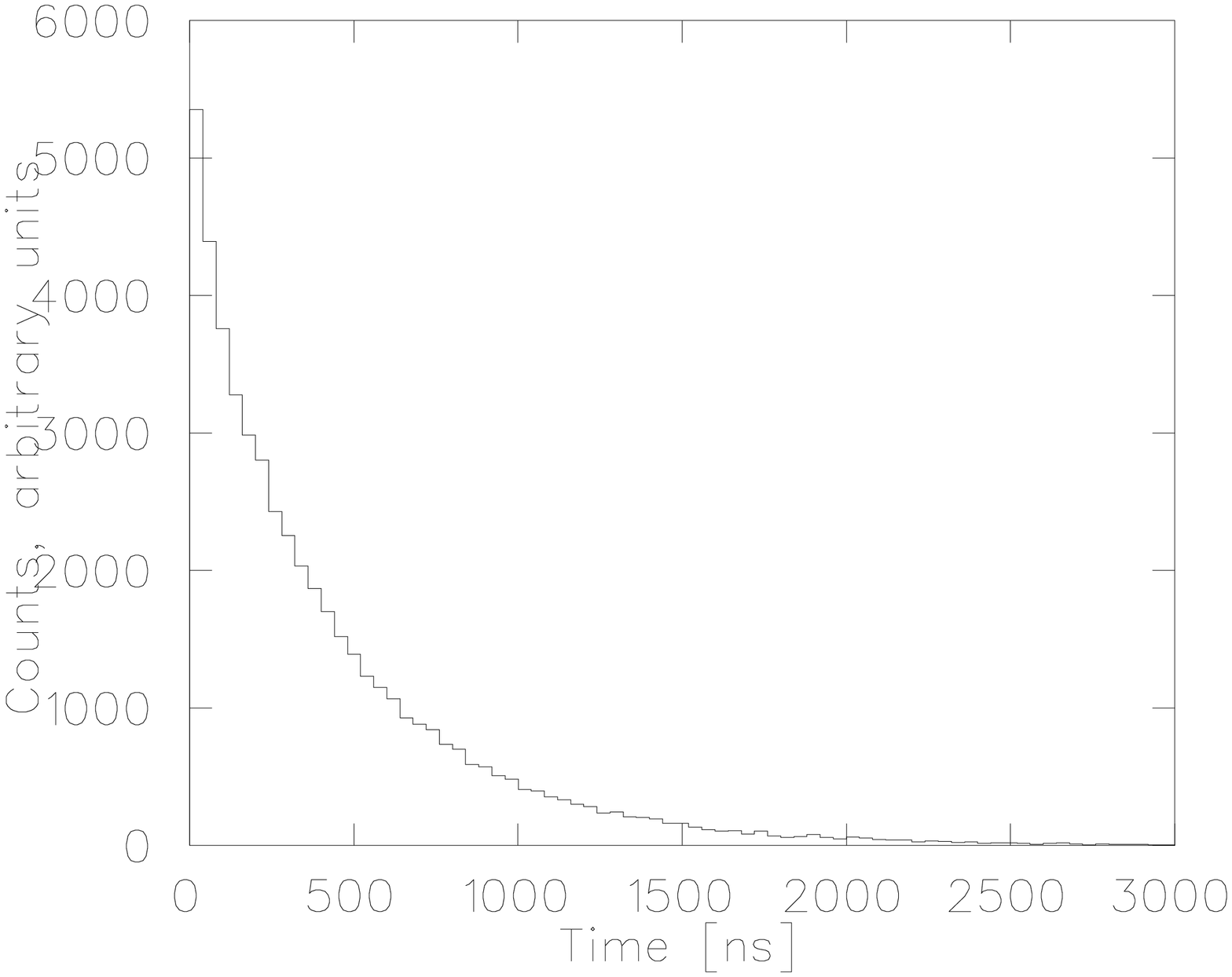}
                \caption{Calculated time distributions 
                of the protons form reactions~(\ref{eq2}a) in
                runs I (left) and II (right).  }
\label{fig:time-cala}
\end{figure*}

\begin{figure*}[htb]
\includegraphics[width=0.35\linewidth,angle=90]{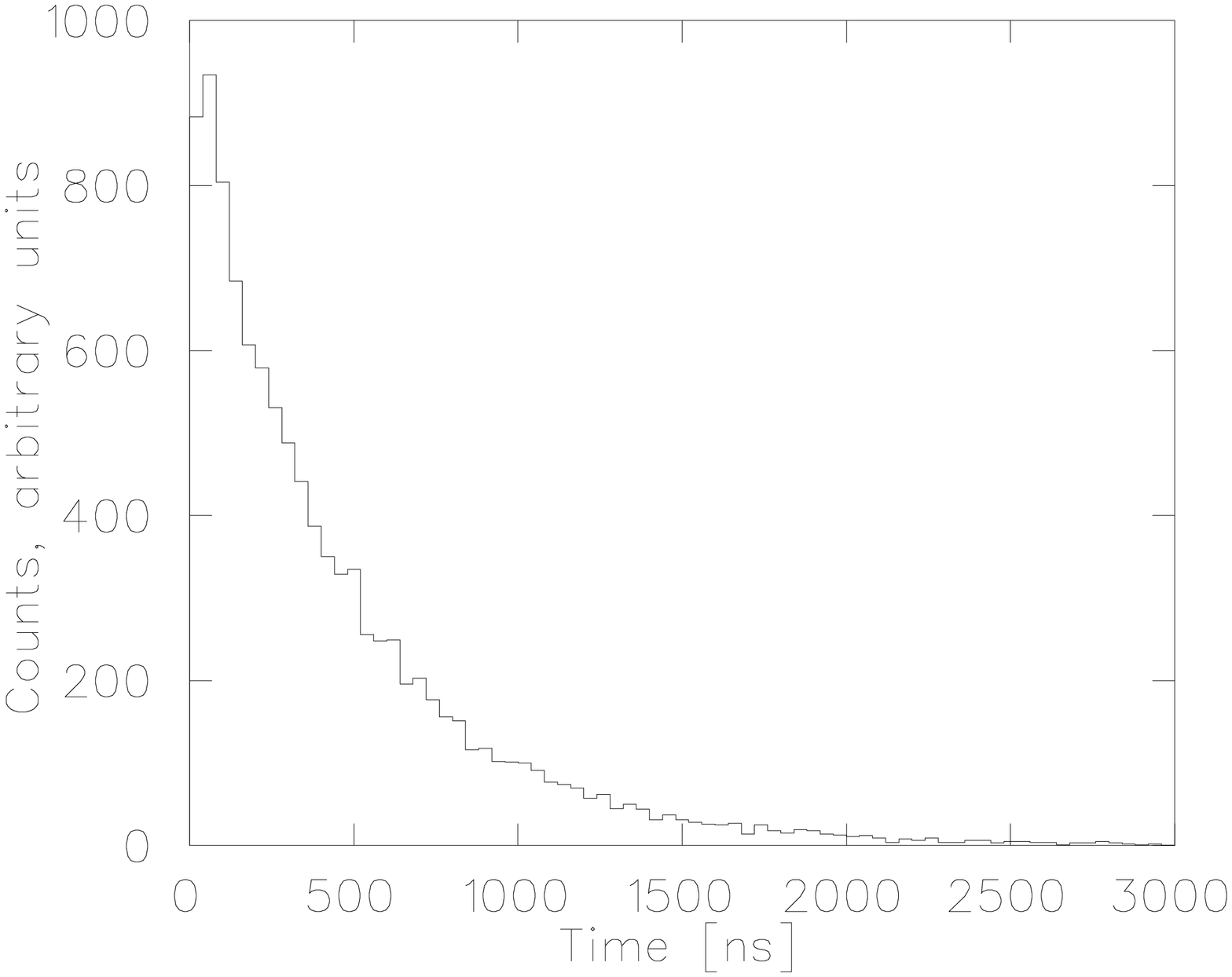}
\includegraphics[width=0.35\linewidth,angle=90]{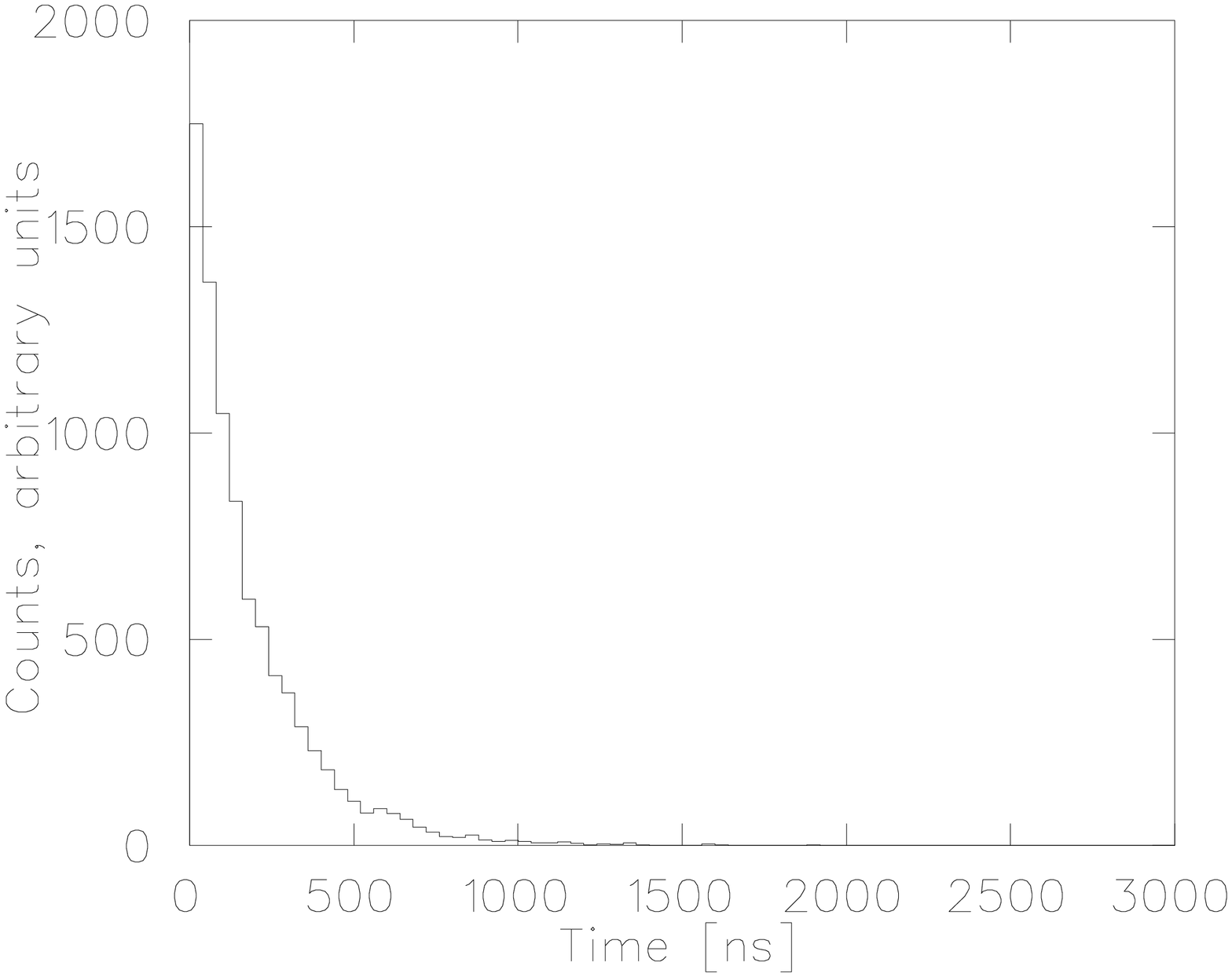}
                \caption{Calculated time distributions
                of the protons form reactions~(\ref{eq31}) in
                runs I (left) and II (right).  }
\label{fig:time-calb}
\end{figure*}

Figures.~\ref{fig:time-cala} and ~\ref{fig:time-calb} show the
calculated time dependencies of the expected yields of protons from
reactions~(\ref{eq2}a) and~(\ref{eq31}) under the conditions of runs I
and II\@.
Thus, there arises a possibility of selecting a time interval of
detection of events by the Si($dE-E$) detectors where the ratio of the
reaction~(\ref{eq2}a) and~(\ref{eq31}) yields is the largest.
This, in turn, makes it possible to suppress the detected background
from reaction~(\ref{eq31}) to a level low enough to meet the
requirement of the experiment on the study of nuclear fusion in the
$d\mu {}^{3}\mathrm{He}$ complex.
Table~\ref{tab:cal-kin} presents the calculated values of some
quantities describing kinetics of muonic processes in the
$\mathrm{D}_2 +{}^{3}\mathrm{He}$ mixture and the process of detecting
protons from reactions~(\ref{eq2}a) and~(\ref{eq31}).

\begin{table}[htb]
\begin{ruledtabular}
      \caption{Calculated values of the quantities describing kinetics
       of muonic processes in the $\mathrm{D}_2 +{}^{3}\mathrm{He}$
       mixture.  The probabilities $W_{{}^3\mathrm{He}}$,
       $W_{d\mu{}^3\mathrm{He}}$, and $W_{d{}^3\mathrm{He}}$ are given
       per muon stop in the gaseous $\mathrm{D}_2 +{}^{3}\mathrm{He}$
       mixture. }
\label{tab:cal-kin}
\begin{tabular}{ccccccccc}
Run & $W_{{}^3\mathrm{He}}$ & $W_{d\mu{}^3\mathrm{He}}$ &
$W_{d{}^3\mathrm{He}}$ & $W_{\mu e}$ & $\varepsilon_p$ &
$\varepsilon_p^{ff}$ & $\eta_p$ & $\eta_p^{ff}$ \\
     & [$10^{-2}$] & [$10^{-1}$] & [$10^{-5}$] & [$10^{-1}$] &
     [$10^{-2}$] & [$10^{-2}$] & [$10^{-8}$] & [$10^{-8}$] \\ \hline
I  & 2.60 & 4.00 & 2.735 & 3.64 & 3.40 & 3.54 & 2.26 & 2.52 \\
II & 2.87 & 5.16 & 2.735 & 2.06 & 3.67 & 3.47 & 2.16 & 2.72 \\
\end{tabular}
\end{ruledtabular}
\end{table}

$W_{{}^3\mathrm{He}}$ is the total probability for the
${}^{3}\mathrm{He}$ formation ($E_{{}^3\mathrm{He}} = 0.8$~MeV) in the
$\mathrm{D}_2 +{}^{3}\mathrm{He}$ mixture, as a result of the fusion
reaction in the $dd\mu$ molecule.
$W_{d\mu{}^3\mathrm{He}}$ is the $d\mu {}^{3}\mathrm{He}$ complex
formation probability and $W_{d{}^3\mathrm{He}}$ is the probability
for $d{}^{3}\mathrm{He}$ fusion in flight, following
reaction~(\ref{eq31}), and $W_{\mu e}$ is the branching ratio of the
muon decay via the $\mu^- \to e^- + \nu_\mu + \bar{\nu}_e$ channel.
$\varepsilon_p$ and $\varepsilon_p^{ff}$ are the detection
efficiencies of one Si($dE-E$) telescope for protons from
reactions (\ref{eq2}a) and~(\ref{eq31}), respectively.
$\eta_p$ and $\eta_p^{ff}$ are the yields of protons from reactions
(\ref{eq2}a) and~(\ref{eq31}) detected by the Si($dE-E$) telescope
per muon stop in the gaseous $\mathrm{D}_2 +{}^{3}\mathrm{He}$
mixture (the value of $d\mu {}^{3}\mathrm{He}$ fusion rate
$\lambda_f = 10^6 \, \mbox{s}^{-1}$ was used for calculation of
$\eta_p$).
There are some noteworthy intermediate results in the calculation of
the detection efficiencies for protons from reactions (\ref{eq2}a)
and~(\ref{eq31}).
Table~\ref{tab:ene-los} presents average energy losses of protons on
their passage through various material in the direction of the
Si($dE-E$) detectors.

\begin{table}[hb]
\begin{ruledtabular}
      \caption{Average energy losses, in MeV, of protons on their
      passage through various materials.}
\label{tab:ene-los}
\begin{tabular}{ccccc}
Run & $\mathrm{D}_2 +{}^{3}\mathrm{He}$ & kapton & Si($dE$) & Si($E$)
\\
    & gas & window & & \\ \hline
I & 1.1 & 0.6 & 3.0 & 10.1 \\
II& 3.5 & 0.7 & 3.7 & 6.9 \\
\end{tabular}
\end{ruledtabular}
\end{table}

\begin{figure*}[t]
\includegraphics[width=0.36\linewidth,angle=90]{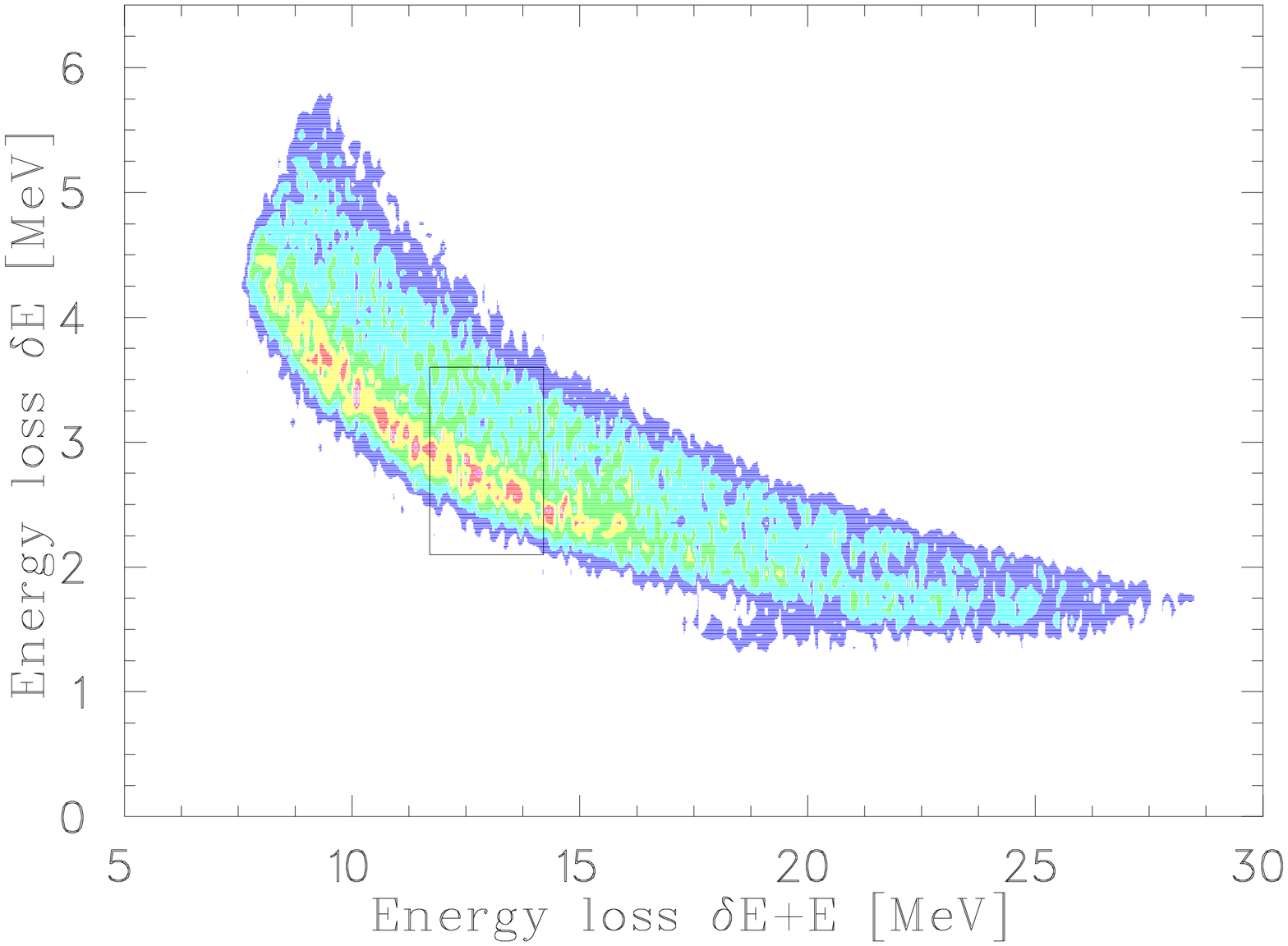}
\hfill
\includegraphics[width=0.36\linewidth,angle=90]{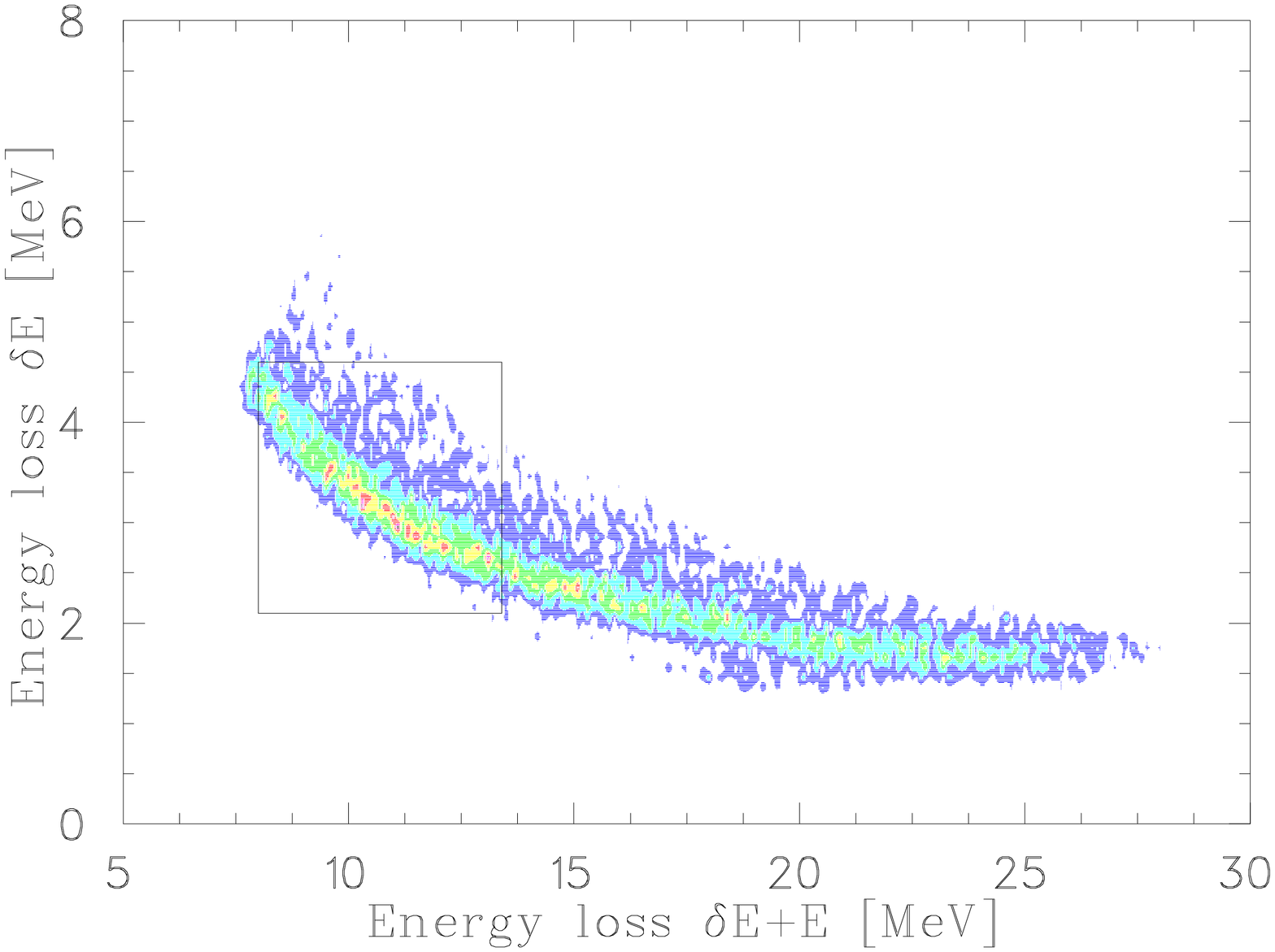}
                \caption{Two-dimensional distributions of events
                detected by the Si($dE-E$) telescopes in runs I (left) and
                II (right).}
\label{fig:si-t1}
\end{figure*}

\begin{figure*}[ht]
\includegraphics[width=0.36\linewidth,angle=90]{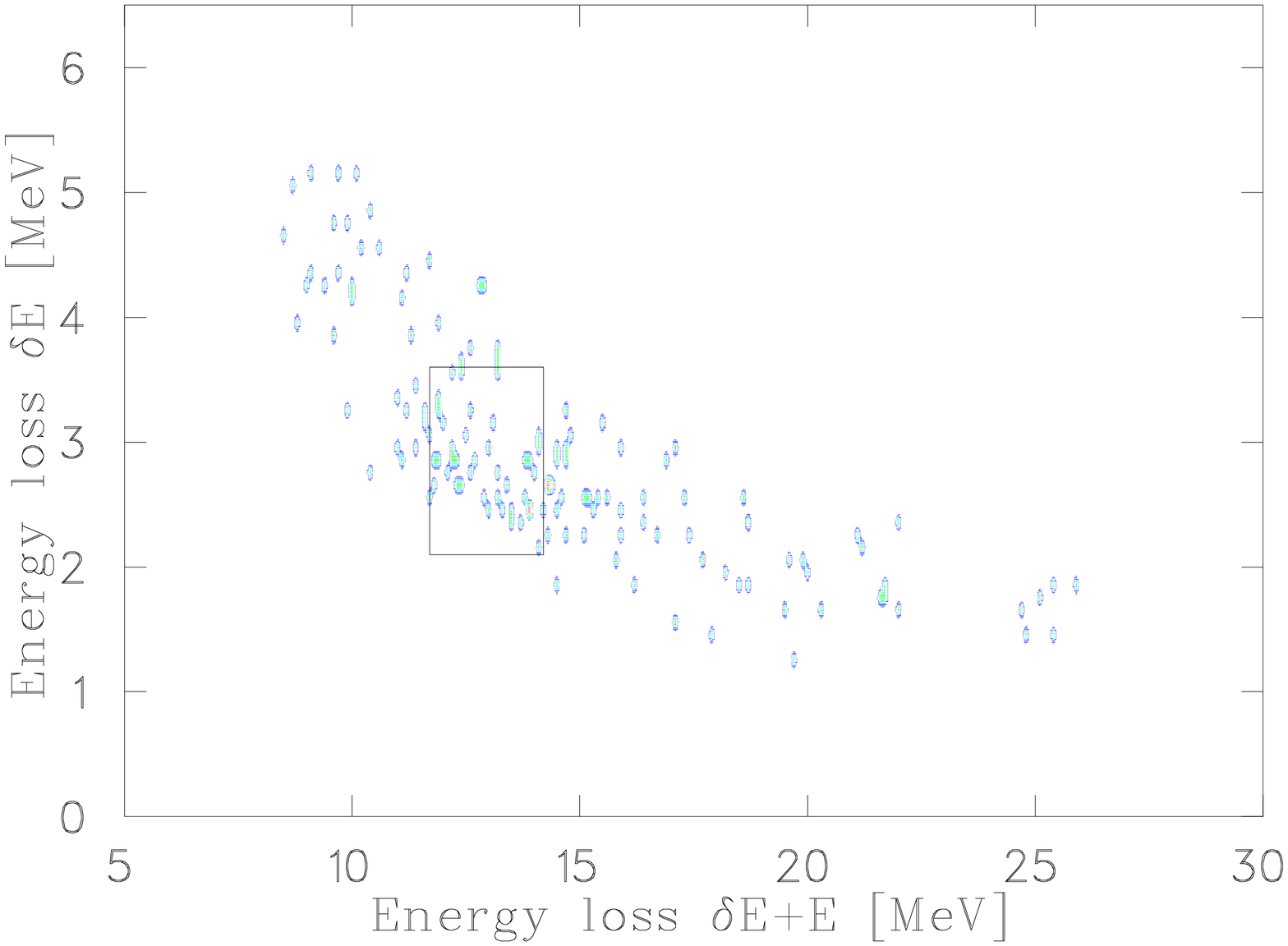}
\hfill
\includegraphics[width=0.36\linewidth,angle=90]{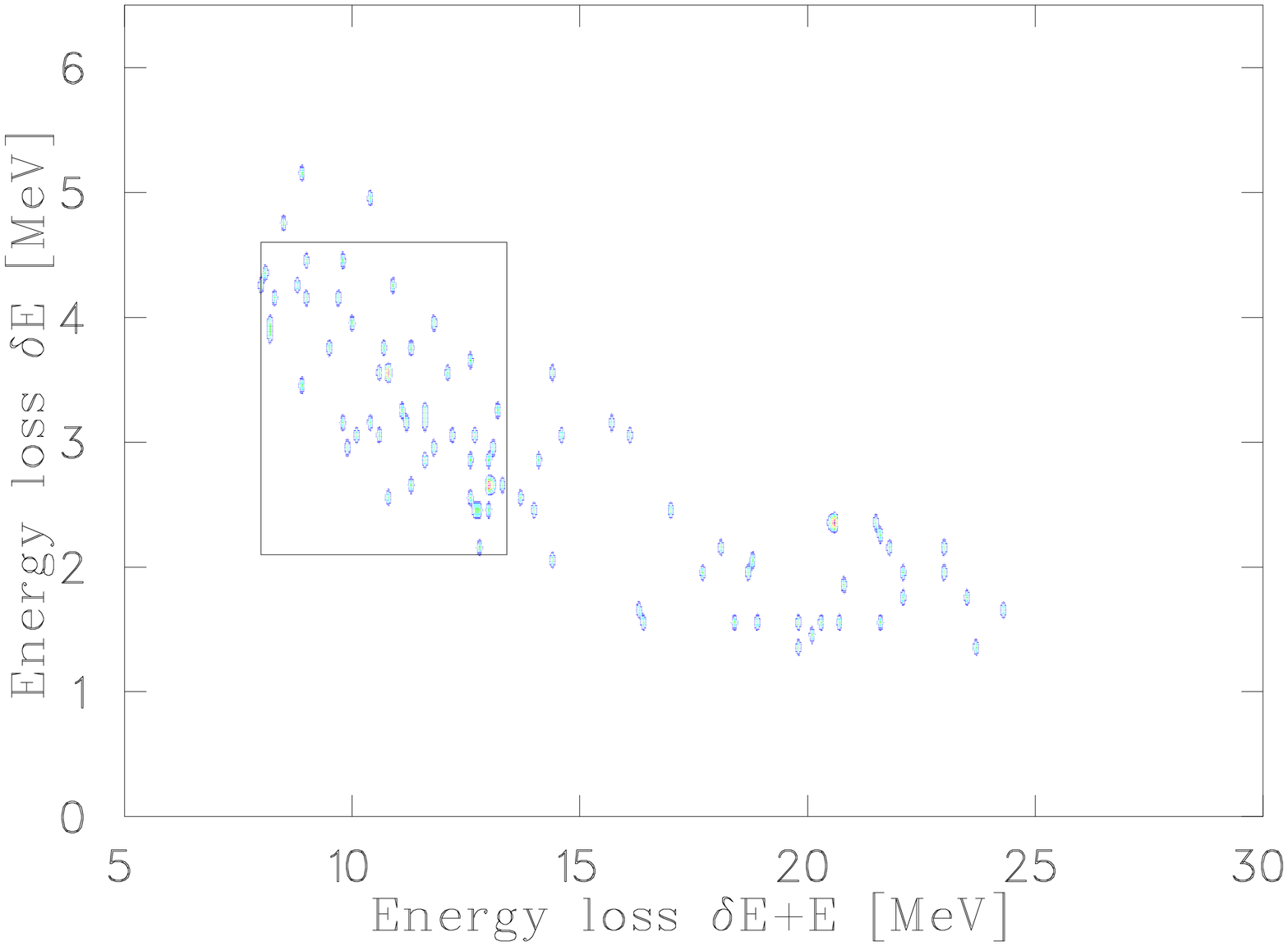}
                \caption{Two-dimensional distributions of events
                detected by the Si($dE-E$) telescopes in runs I (left)
                and II (right) with the del-$e$ coincidence.}
\label{fig:si-t2}
\end{figure*}

Figures~\ref{fig:si-t1} show the two-dimensional distributions of
events detected by the Si($dE-E$) detectors without coincidences with
muon decay electrons in runs I and II\@.
The x--axis represents the energy losses in the thin Si($dE$) counters
and the y--axis shows the total energies losses by the particle in
both the Si($dE$) and Si($E$) detectors connected in coincidence.
The distributions of events in Figs.~\ref{fig:si-t1} correspond to the
detection of protons arising both from reactions~(\ref{eq2}a)
and~(\ref{eq31}) and from the background reactions such as
\begin{eqnarray}
  \label{eq32}
       \mu + {}^3\mathrm{He} & \to & p + 2 n + \nu_\mu \nonumber \\
       \mu + \mathrm{Al} &\to \mathrm{Na}^* & + p + n + \nu_\mu  \nonumber \\
                         &                  & + p + \nu_\mu \nonumber\\
                         &                  & + p + 2n + \nu_\mu \\
       \mu + \mathrm{Fe} &\to \mathrm{Cr}^* &+ p + n + \nu_\mu \nonumber  \\
                         &                  & + p + \nu_\mu \nonumber\\
                         &                  & + p + 2n + \nu_\mu \, . \nonumber
\end{eqnarray}
In addition, the background which is not correlated with muon stops in
the target (background of accidental coincidences) contributes to
these distributions.

Figures~\ref{fig:si-t2} show the two-dimensional Si($dE - (E + dE)$)
distributions obtained in coincidences with muon decay electrons.
As seen, the use of the del-$e$ criterion leads to an appreciably
reduction of the background, which in turn makes it possible to
identify a rather weak effect against the intensive background signal.
To suppress muon decay electrons in the Si($dE-E$) telescope,
provision was made in the electronic logic of the experiment to
connect each of the electron detectors in anti-coincidence with the
corresponding Si($dE-E$) telescope.
The choice of optimum criteria in the analysis of the data from the
Si($dE-E$) telescopes was reduced to the determination of the
boundaries and widths of the time and energy intervals where the
background is substantially suppressed in absolute value and the
effect-to-background ratio is the best.
To this end the two-dimensional Si($dE - (dE + E)$) distributions
corresponding to the detection of protons were simulated by MC method
for runs I and II\@.
On the basis of these distributions boundaries were determined for the
energy interval of protons from reaction (\ref{eq2}a) where the loss
of the ``useful'' event statistics collected by the Si telescope would
be insignificant.

\begin{figure*}[ht]
\includegraphics[width=0.45\linewidth]{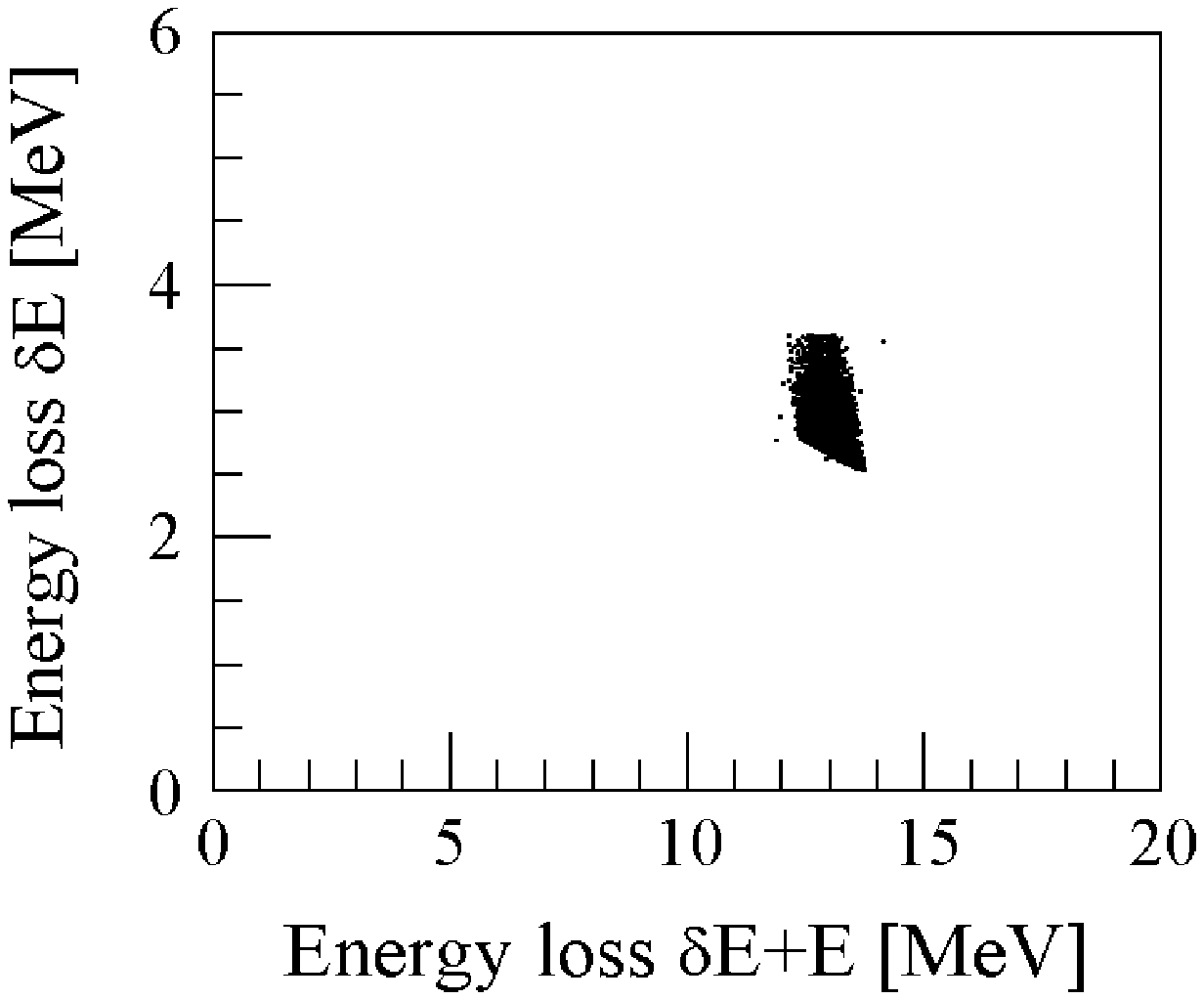}
\includegraphics[width=0.45\linewidth]{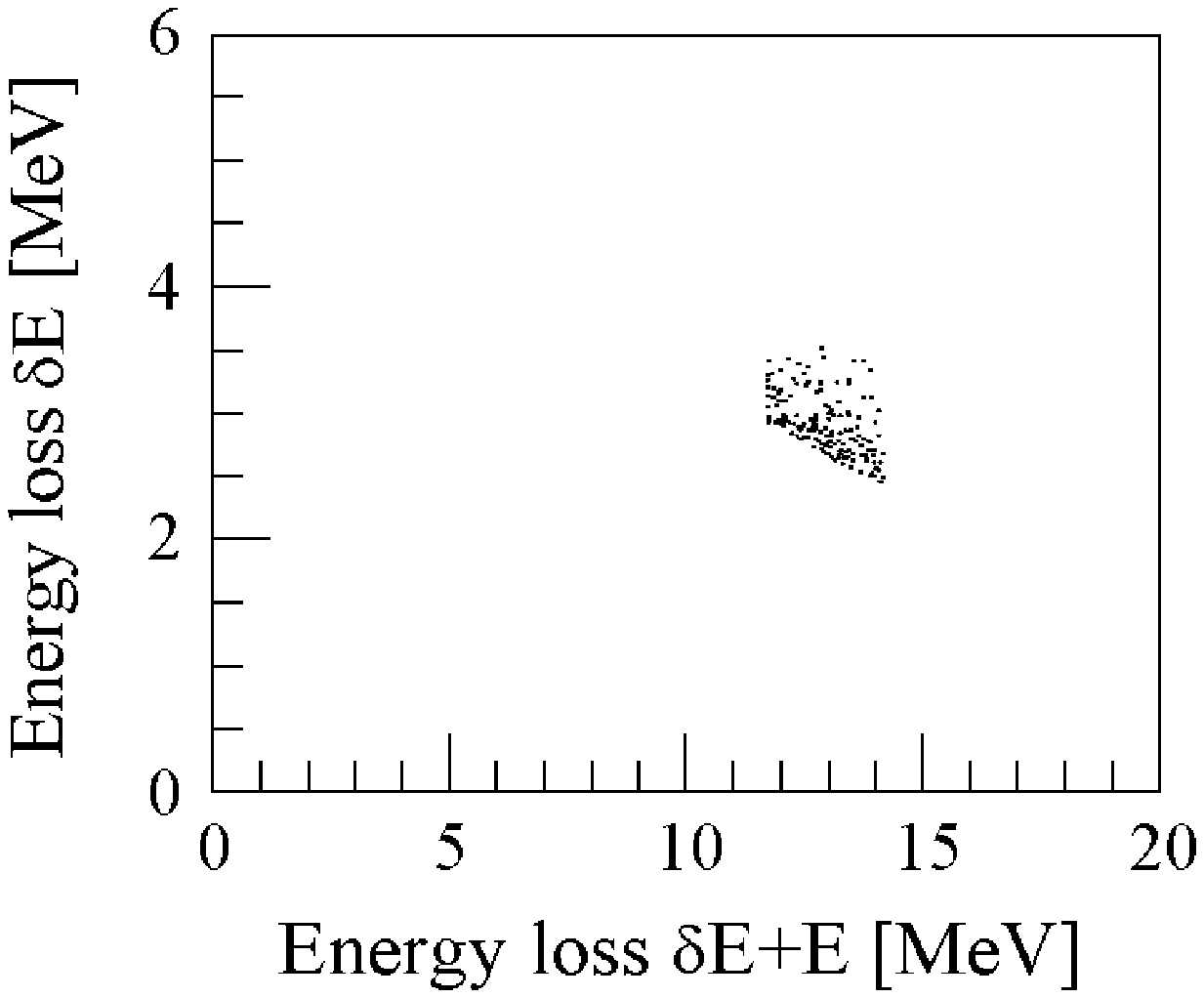}
                \caption{Two-dimensional distributions of
                Si($dE-(dE+E)$) events obtained in run I by the Monte
                Carlo method and corresponding to detection of protons
                from reactions~(\ref{eq2}a) (left) and~(\ref{eq31})
                (right) in the time interval $\Delta t_\mathrm{Si}$.}
\label{fig:si-mc-t1}
\end{figure*}

\begin{figure*}[ht]
\includegraphics[width=0.45\linewidth]{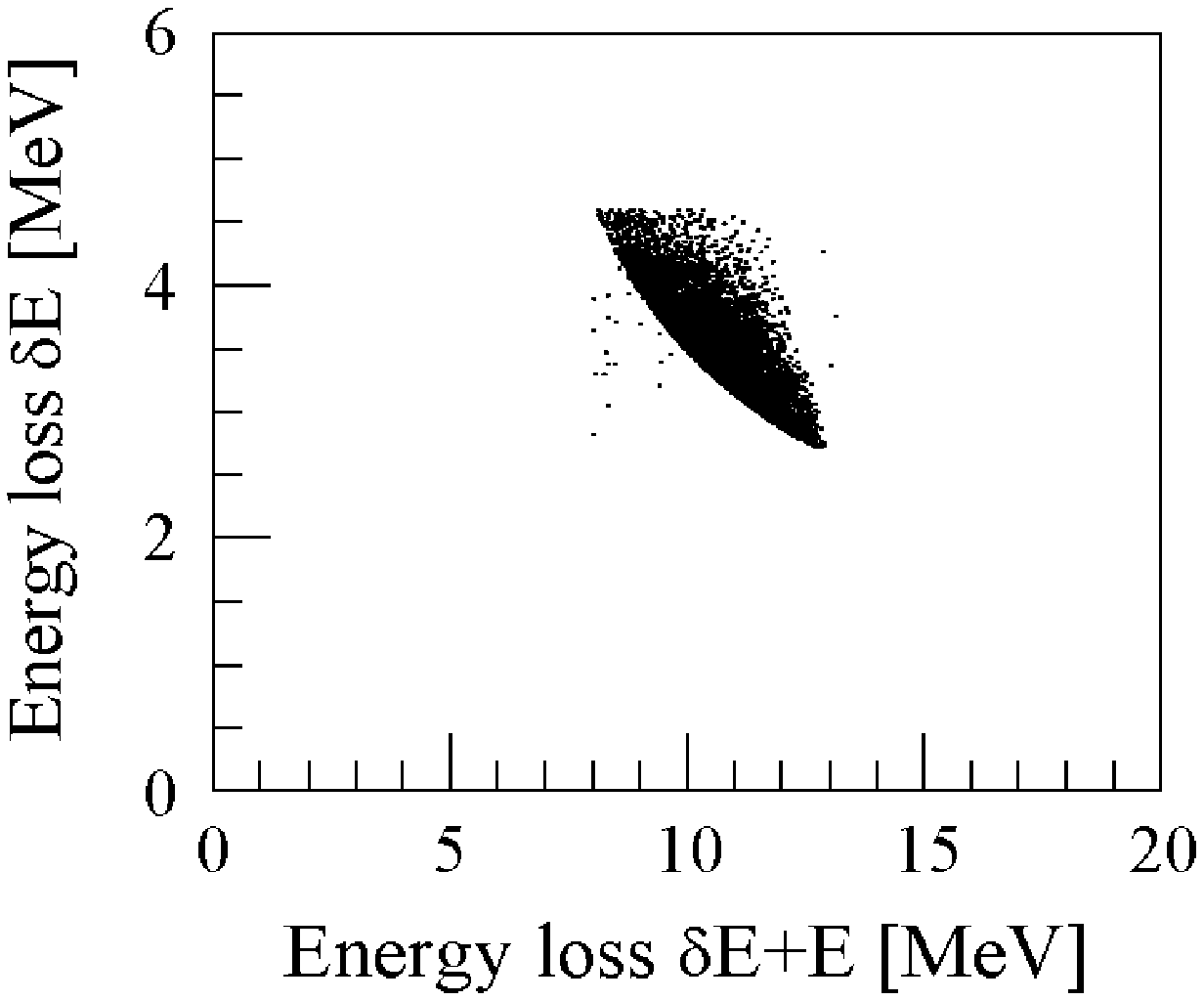}
\includegraphics[width=0.45\linewidth]{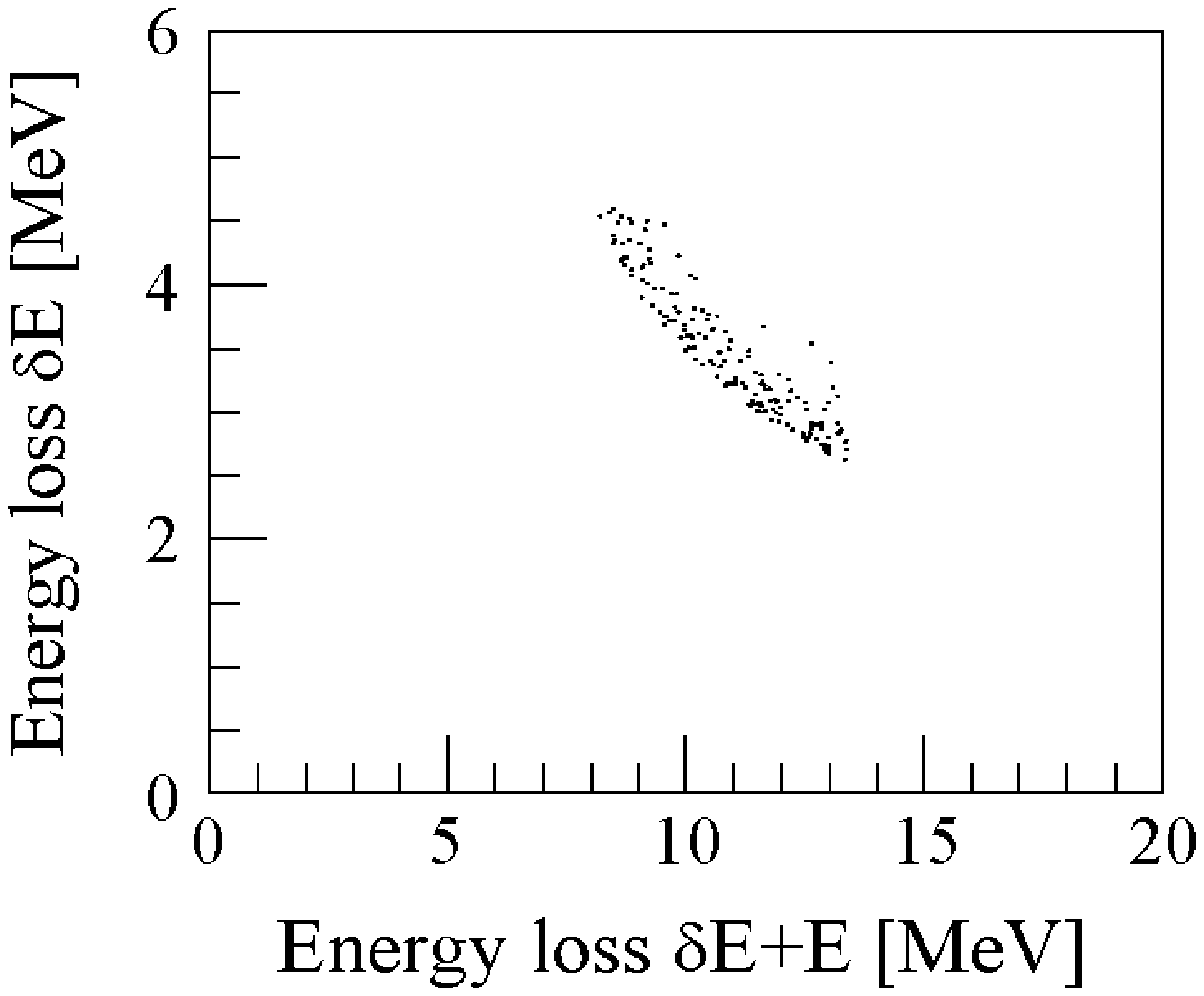}
                \caption{Two-dimensional distributions of
                Si($dE-(dE+E)$) events obtained in run II by the Monte
                Carlo method and corresponding to detection of protons
                from reactions~(\ref{eq2}a) (left) and~(\ref{eq31})
                (right) within the time interval $\Delta
                t_\mathrm{Si}$.}
\label{fig:si-mc-t2}
\end{figure*}

Figures~\ref{fig:si-mc-t1} and \ref{fig:si-mc-t2} show the
two-dimensional Si($dE - (dE + E)$) distributions corresponding to the
proton detection which were simulated by the MC method for runs I and
II\@.
Based on these distributions, we chose some particular proton energy
intervals named $\Delta E_\Sigma$ when considering the total energy
deposited and $\delta E$ when looking only at the Si($dE$) detector
(see Table~\ref{tab:time-factor}) for further analysis.
The regions of events corresponding to the intervals $\delta E$ and
$\Delta E_\Sigma$ are shown in the form of rectangles on the two-
dimensional distributions presented on the Fig.~\ref{fig:si-t1}
and~\ref{fig:si-t2}.

It is noteworthy that the proton detection efficiencies given in
Table~\ref{tab:cal-kin} correspond to these chosen proton energy
intervals for runs I and II\@.
The next step in the data analysis was to choose a particular times
interval of detection of events by the Si($dE-E$) telescope.
Figures~\ref{fig:si-mc-t1} and \ref{fig:si-mc-t2} show the simulated
time distributions of protons corresponding to the chosen energy loss
intervals $\delta E$, for the energy loss in the Si($dE$) detector and
$\Delta E_\Sigma = E + \delta E$ the energy loss in both silicon
detector.
For the chosen proton energy intervals Table~\ref{tab:time-factor}
presents the statistics suppression factors corresponding to different
initial time,$t_{thr}$ (with respect to the instant of the muon stop
in the target) of the time intervals of detection of proton events.
These factors correspond to the $\varepsilon_Y$ value in
Eq.~(\ref{equ:pyieldt}).
The data in Table~\ref{tab:time-factor} are derived from time
dependencies of the yields of protons from reactions~(\ref{eq2}a)
and~(\ref{eq31}) (see Figs.~\ref{fig:time-cala}
and~\ref{fig:time-calb}).

\begin{table*}[ht]
\begin{ruledtabular}
      \caption{Time factors for reaction~(\ref{eq2}a) and reaction
      (\ref{eq31}) for the chosen intervals of energies of protons
      detected by the Si($dE-E$) telescopes and detection beginning
      time, $t_{thr}$.}
\label{tab:time-factor}
\begin{tabular}{ccc|ccccc|ccccc}
Run & $\Delta E_\Sigma$ & $\delta E$ &
\multicolumn{5}{c}{Reaction~(\ref{eq2}a)} &
\multicolumn{5}{c}{Reaction~(\ref{eq31}) } \\
& [MeV] & [MeV] & \multicolumn{5}{c}{$t_{thr},\
\mu\mathrm{s}^{-1}$}&
\multicolumn{5}{c}{$t_{thr},\ \mu\mathrm{s}^{-1}$}\\
\hline
& & & 0.0 & 0.2 & 0.4 & 0.7 & 0.9 & 0.0 & 0.2 & 0.4 & 0.7 & 0.9
\\ \hline
I & [$0-\infty$] & [$0-\infty$] & 0.911 & 0.684 & 0.524 & 0.350 &
0.264 &
0.989 & 0.599 & 0.388 & 0.198 & 0.131 \\
  & [$11.7-14.2$] & [$2.1-3.6$] & 0.878 & 0.659 & 0.505 & 0.337 &
  0.254 & 0.438 & 0.263 & 0.171 & 0.090 & 0.058 \\
II& [$0-\infty$]  & [$0-\infty$] & 0.934 & 0.543 & 0.316 & 0.129 &
0.059 &
0.996 & 0.333 & 0.114 & 0.025 & 0.009 \\
  & [$8.0-13.4$] & [$2.1-4.6$]  & 0.904 & 0.525 & 0.306 & 0.125 &
  0.057 & 0.752 & 0.252 & 0.084 & 0.018 & 0.006 \\
\end{tabular}
\end{ruledtabular}
\end{table*}

\begin{figure*}[ht]
\includegraphics[width=0.36\linewidth,angle=90]{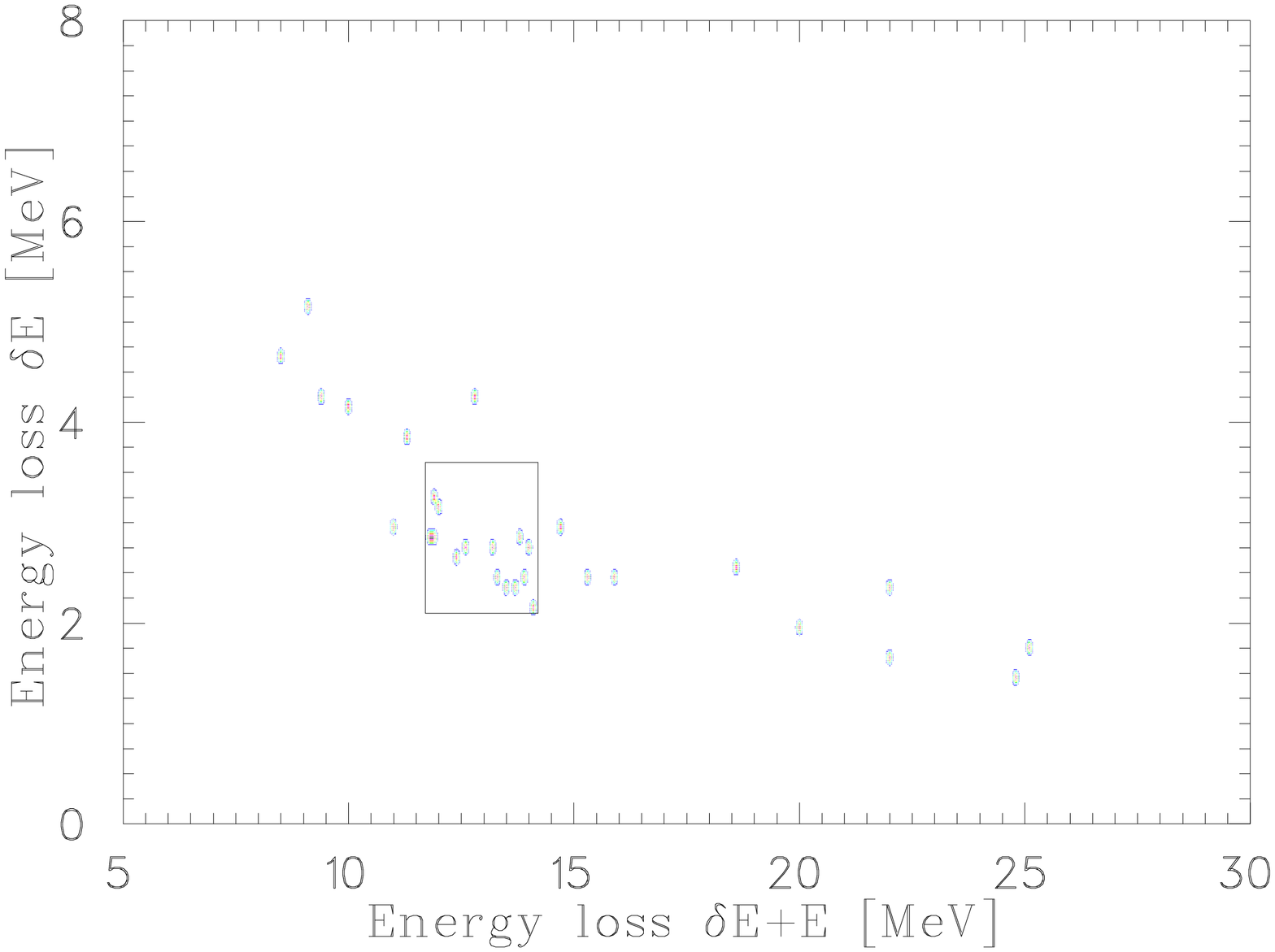}
\hfill
\includegraphics[width=0.36\linewidth,angle=90]{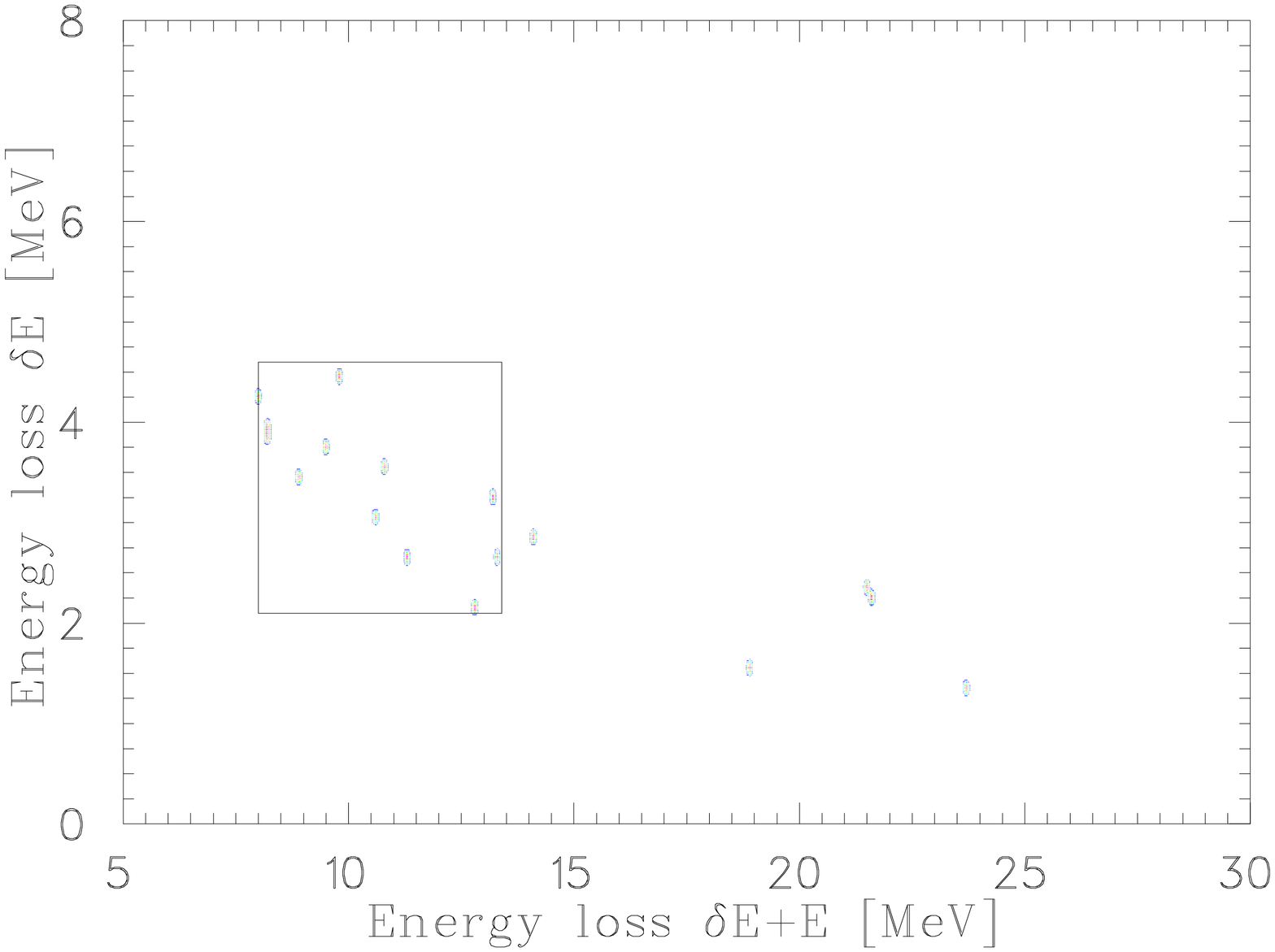}
                \caption{Two-dimensional distributions of events
                detected by the Si($dE-E$) telescopes in runs I (left)
                and II (right) with the del-$e$ coincidence and the
                time interval $\Delta t_\mathrm{Si}$ as define in
                Eq.~(\ref{eq:tsi}).  }
\label{fig:si-t3}
\end{figure*}

According to the data given in Table~\ref{tab:time-factor}, we took
the following time intervals $\Delta t_\mathrm{Si}$ (with
$t_\mathrm{Si}$ the time for the Si signal to appear) for analyzing
the events
\begin{eqnarray}
  \label{eq:tsi}
      \Delta t_\mathrm{Si} \, \mbox{(run I)}: \qquad 0.7 \leq &
      t_\mathrm{Si} & \leq 2.2 \, \mu \mbox{s} \nonumber \\
      \Delta t_\mathrm{Si} \, \mbox{(run II)}: \qquad 0.4 \leq &
      t_\mathrm{Si} & \leq 1.2 \, \mu \mbox{s} \, .
\end{eqnarray}
Figures~\ref{fig:si-t3} display the two-dimensional distributions of
Si($dE-E$) events obtained in coincidences with muon decay electrons
in runs I and II with this time criteria imposed.
With these time intervals $\Delta t_\mathrm{Si}$ and the proton energy
loss $\Delta E_\Sigma$ and $\delta E$ intervals, the statistics
collection suppression factors for events from reactions (\ref{eq2}a)
and~(\ref{eq31}) are
\begin{eqnarray}
  \label{eq33}
  k_{d \mu {}^3\mathrm{He}} = 2.9, &  k_{d {}^3\mathrm{He}} & =
  11.2,
  \qquad   \mbox{Run I,} \nonumber \\
  k_{d \mu {}^3\mathrm{He}} = 3.2, & k_{d {}^3\mathrm{He}} & =
  12.1,
  \qquad \mbox{Run II.} \nonumber \\
\end{eqnarray}

Another stage of the data analysis was the determination of the number
of events detected by the Si($dE-E$) telescopes in runs I and II under
the following criteria
\begin{itemize}
  \item[(i)] the coincidence of signals from the Si telescopes and
            electron detectors in the time interval $0.2 < (t_e -
            t_\mathrm{Si}) < 5.5 \,\mu$s ($t_e$ is the time when the E
            detector signal appear). Such a requirement add the
            efficiency factor $\varepsilon_t = 0.83$ when determining
            the rates.
  \item[(ii)] the total energy release in the Si($dE$) detector is
            $\delta E$ as given in Table~\ref{tab:time-factor}. This
            particular $\delta E$ interval will be called $\delta
            E$. For the thin and thick Si detector together, we
            choose the smallest interval, namely $\Delta E_\Sigma =
            [11.7-14.2]$~MeV for run I and $\Delta E_\Sigma =
            [8.0-13.4]$~MeV for run II\@.
  \item[(iii)] the time when the signal from the Si telescope appears
            falls in the $\Delta t_\mathrm{Si}$ intervals.
\end{itemize}
Table~\ref{tab:number} presents the numbers of events $N_p$ detected
in runs I and II under the above mentioned criteria.

\begin{table}[ht]
\begin{ruledtabular}
      \caption{Numbers of detected events, $N_p$ and $N_p^{ff}$ for
      the chosen $\delta E$ and $\Delta E_\Sigma$ intervals,
      taking into account the time intervals $t_e - t_\mathrm{Si}$,
      and $\Delta t_\mathrm{Si}$. Also the accidental coincidence
      background, $N_p^{acc}$, as well as the total background,
      $N_p^{bckg}$.}
\label{tab:number}
\begin{tabular}{cccccc}
Run & $N_p$ & $N_p^{ff}$ & $\eta_{Si-E}$ & $N_p^{acc}$ & $N_p^{bckg}$
\\
& & & [$10^{-4}$] & & \\ \hline
I & 14 & 3.8(2) & 4.2(9) & 2.5(5) & 6.3(6) \\
II & 11 & 2.4(1) & 2.4 (11) & 1.1(5) & 3.5(5) \\
\end{tabular}
\end{ruledtabular}
\end{table}

The contribution of the background events, $N_p^{ff}$, given in
Table~\ref{tab:number} from the reaction~(\ref{eq31}) is found in the
following way.
The expected number of detected protons from reaction~(\ref{eq31}) in
runs I and II is calculated by
\begin{equation}
\label{eq34}
      N_p^{ff} = \frac{N_\mu a_{D/He} W_{{}^3\mathrm{He}}
      W_{d{}^3\mathrm{He}} \varepsilon^{ff}_p N_\mathrm{Si} \varepsilon_e
      \varepsilon_t}{k_{d {}^3\mathrm{He}}} \, .
\end{equation}
$N_\mathrm{Si}$ is the number of Si($dE-E$) telescopes and $1/{k_{d
{}^3\mathrm{He}}}$ is the factor of background suppression by imposing
the criteria (ii) and (iii).
Using the values of $a_{D/He}$ and $N_\mu$ measured in runs I and II,
the calculated values of $W_{^3He}$, $W_{d^3He}$, $\varepsilon^f_p$,
$N_\mathrm{Si}$, $k_{d {}^3\mathrm{He}}$, $\varepsilon_t$, and
Eq.~(\ref{eq34}), we obtained $N_p^{ff}$, which is given in
Table~\ref{tab:number}.
Errors of the calculated $N_p^{ff}$ arose from the inaccurate
dependence of the cross sections $\sigma_{d{}^3\mathrm{He}}$ for the
$d{}^{3}\mathrm{He}$ reaction in flight on the ${}^{3}\mathrm{He}$
deuteron collision energy and from the errors in the calculations of
the detection efficiency of the Si telescopes for protons from
reaction~(\ref{eq31}).
These errors were found by substituting various experimental
$\sigma_{d^3He}(E_{d^3He})$
dependencies~\cite{white97,kunzx55,kljuc56,allre52,argox52,freie54}.
into the program for Monte Carlo calculation of the in-flight
$d{}^{3}\mathrm{He}$ fusion probability $W_{d{}^3\mathrm{He}}$.

Now it is necessary to find the level of the accidental coincidence
background by analyzing the experimental data from runs I and II\@.
To this end the two-dimensional distribution of events detected by the
Si($dE-E$) telescopes was divided into three regions which did not
include the separated region of events belonging to the
process~(\ref{eq2}a).
Considering the boundaries of the intervals $\delta E$ and
$\Delta E_\Sigma$ of energy losses of the protons from
reaction~(\ref{eq2}a) we used three regions, A,B, and C, of the
two-dimensional $\delta E - \Delta E_\Sigma$ distributions for
determining the background level.
The regions are given in Table~\ref{tab:divi}.

\begin{figure}[ht]
\includegraphics[width=0.7\linewidth,angle=90]{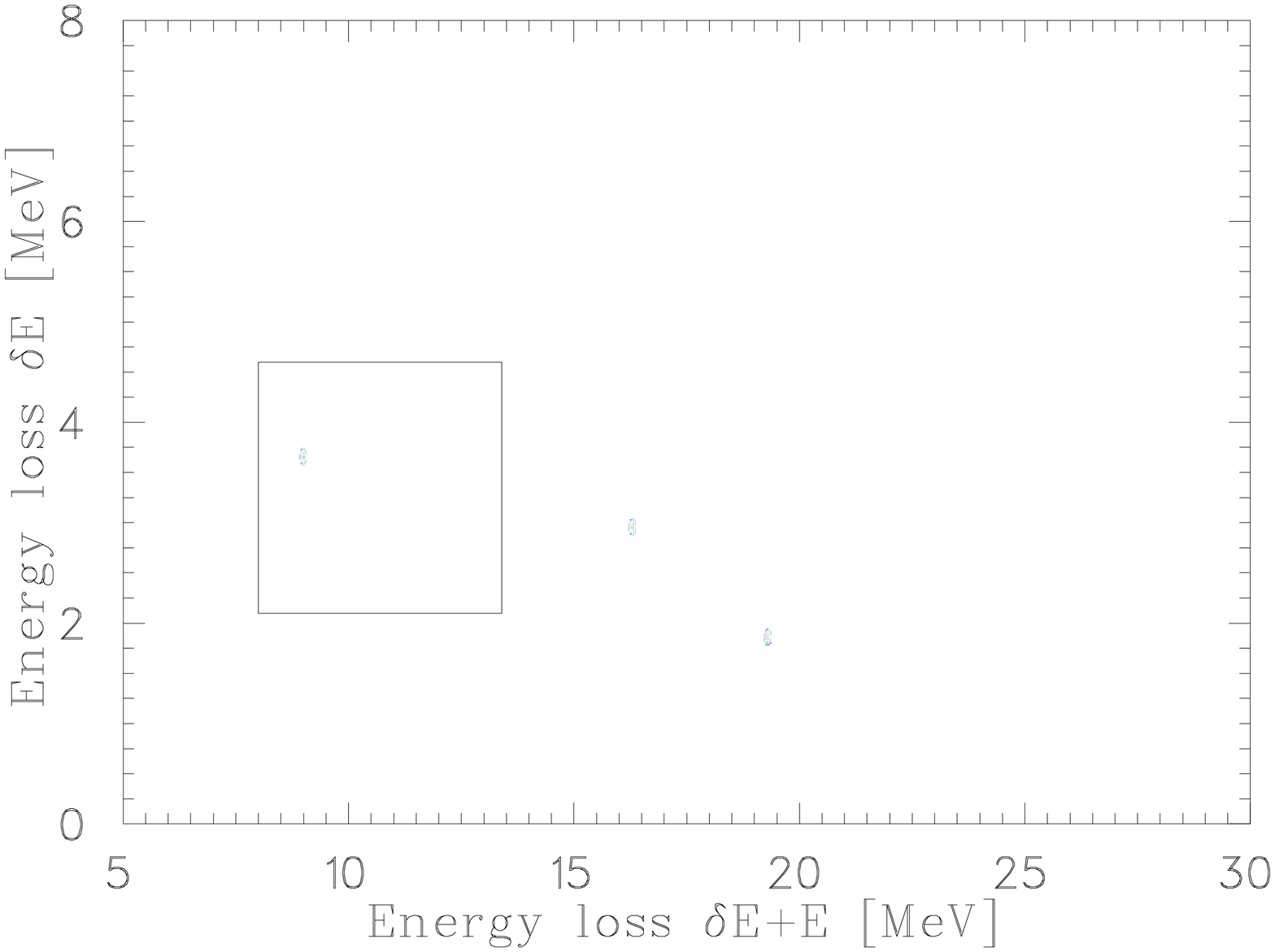}
                \caption{Two-dimensional distributions of events
                detected by the Si($dE-E$) telescopes in a run with
                pure deuterium with the del-$e$ coincidences and within
                the $\Delta t_\mathrm{Si}$ interval. The rectangle is
                the region corresponding to the energy intervals
                $\delta E$ and $\Delta E_\Sigma$ for the run with the
                $\mathrm{D}_2 +{}^{3}\mathrm{He}$ mixture at $\varphi
                = 0.168$.}
\label{fig:si-d2h}
\end{figure}

\begin{figure}[ht]
\includegraphics[width=0.7\linewidth,angle=90]{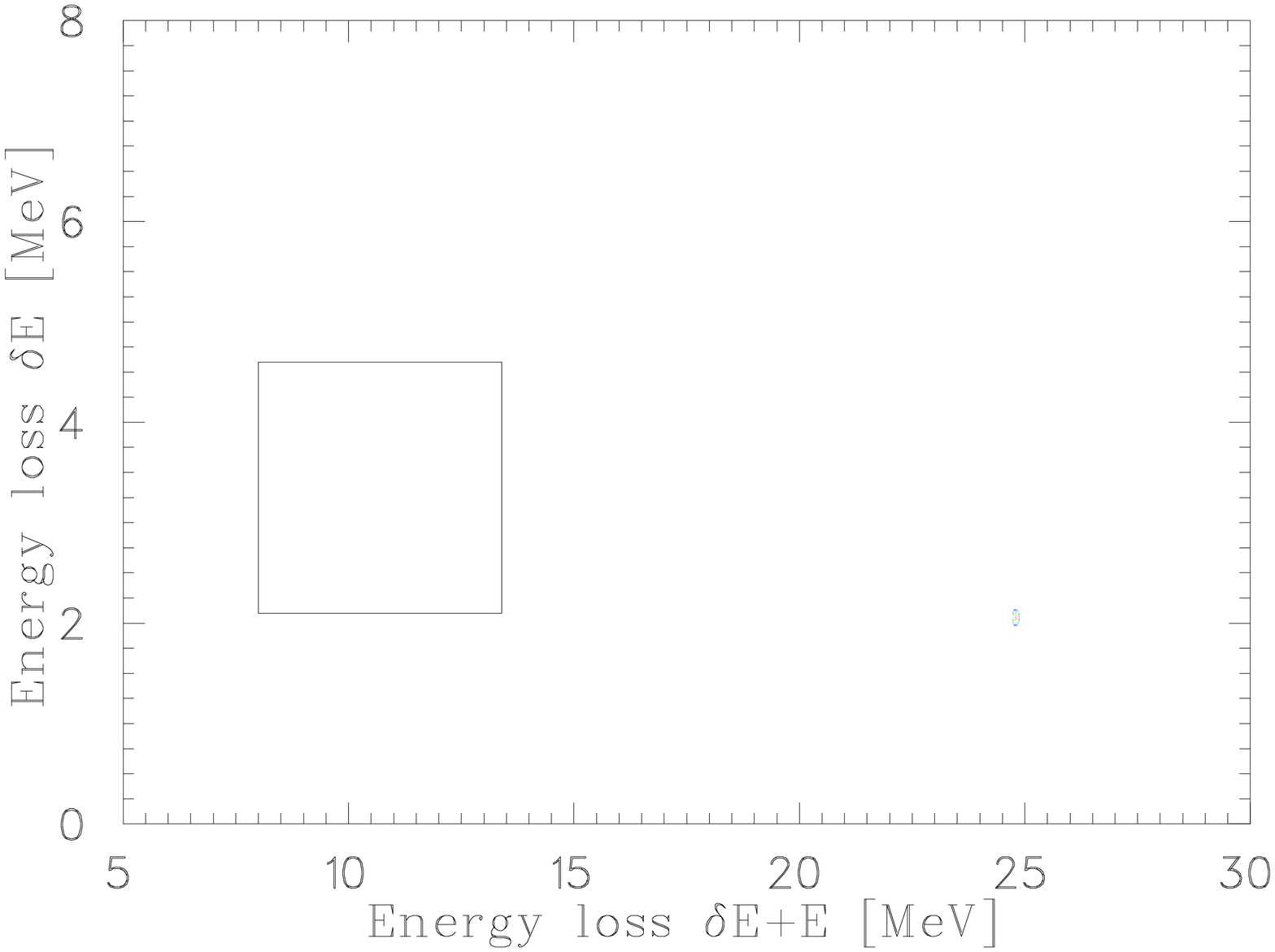}
                \caption{Two-dimensional distributions of events
                detected by the Si($dE-E$) telescopes in the pure
                ${}^{4}\mathrm{He}$ with the del-$e$ coincidences and
                within the $\Delta t_\mathrm{Si}$ interval. The
                rectangle is region corresponding to the energy
                intervals $\delta E$ and $\Delta E_\Sigma$ for the run
                with the $\mathrm{D}_2 +{}^{3}\mathrm{He}$ mixture at
                $\varphi = 0.168$.  }
\label{fig:si-he4}
\end{figure}

\begin{figure}[htb]
\includegraphics[width=0.7\linewidth,angle=90]{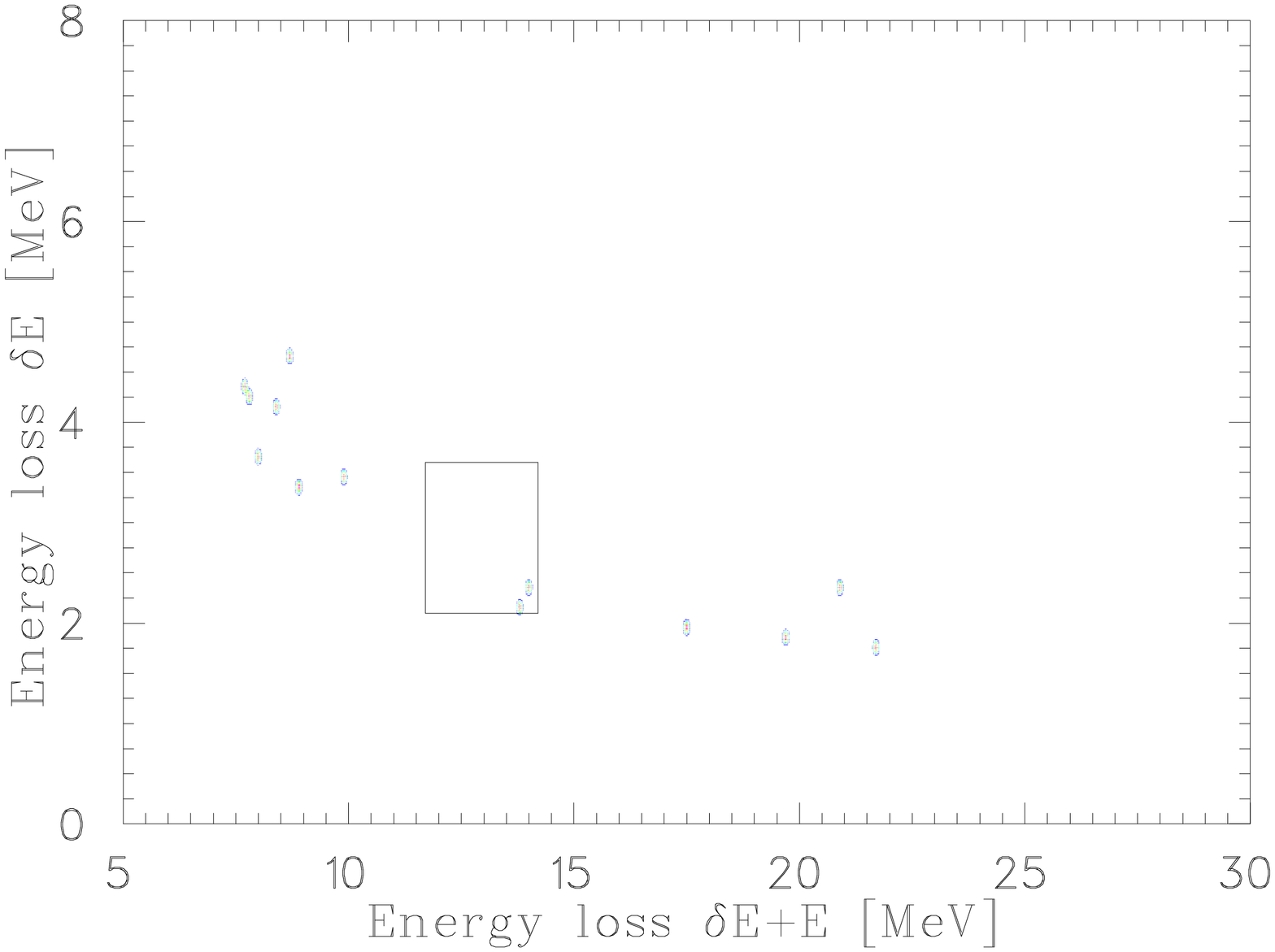}
                \caption{Two-dimensional distributions of events
                detected by the Si($dE-E$) telescopes in the run with
                pure ${}^{3}\mathrm{He}$ with the del-$e$ coincidence
                and within the $\Delta t_\mathrm{Si}$ interval. The
                rectangle is region corresponding to the energy
                intervals $\delta E$ and $\Delta E_\Sigma$ for the run
                with the $\mathrm{D}_2 +{}^{3}\mathrm{He}$ mixture at
                $\varphi = 0.0585$.  }
\label{fig:si-he3}
\end{figure}

The level $N_p^{acc}$ of the background of the accidental coincidences
of signals from the Si($dE-E$) telescopes and the electron detectors
for the given three region of the two-dimensional $\delta E - \Delta
E_\Sigma$ distributions and the corresponding suppression factor of
the accidental background in the Si telescopes, $\eta_{Si-E}$, are
defined as
\begin{equation}
\label{eq35}
      N_p^{acc} = N^f_\mathrm{Si} \eta_{Si-E},
\end{equation}
\begin{equation}
\label{eq36}
      \eta_{Si-E} = \frac{\sum\limits_i N^i_{Si-E}}{\sum\limits_i
      N^i_{Si}},
\end{equation}
where $N^f_\mathrm{Si}$ is the number of events detected by the
three Si($dE-E$) telescopes and belonging to the selected
$(\delta E - \Delta E_\Sigma)$ region of detection of protons
from reaction (\ref{eq2}a).
$N^i_{Si-E}$ and $N^i_{Si}$ are the numbers of events detected by
the $i$th Si($dE-E$) telescope with and without del-$e$ coincidences
and belonging to the other $(\delta E - \Delta E_\Sigma)$ intervals.
Note that the degree of suppression of the accidental coincidence
background was determined not only by averaging the data obtained with
the $\mathrm{D}_2 +{}^{3}\mathrm{He}$ mixture but also in additional
experiments with the targets filled with pure ${}^{4}\mathrm{He}$,
$\mathrm{D}_2$, and ${}^{3}\mathrm{He}$ whose densities were $\varphi
\approx 0.17$, $\varphi \approx 0.09$, and $\varphi \approx
0.035$,respectively.
This guaranteed an identical ratio of stops in the target walls and in
the gas in the experiments with ${}^{4}\mathrm{He}$, $\mathrm{D}_2$,
and the $\mathrm{D}_2 +{}^{3}\mathrm{He}$ mixture ($\varphi = 0.168$)
and in the experiments with the $\mathrm{D}_2 +{}^{3}\mathrm{He}$
mixture ($\varphi = 0.0585$) and ${}^{3}\mathrm{He}$.
Figures~\ref{fig:si-d2h},~\ref{fig:si-he4}, and~\ref{fig:si-he3}
display the two-dimensional distributions of background events
detected by the Si($dE-E$) telescopes in the experiments with
${}^{4}\mathrm{He}$, $\mathrm{D}_2$, and ${}^{3}\mathrm{He}$.

\begin{table}[hb]
\begin{ruledtabular}
      \caption{The three regions of division of two-dimensional
      ($\delta E - \Delta E_\Sigma$) distributions. All energies are
      given in MeV\@.}
\label{tab:divi}
\begin{tabular}{ccccccc}
 & \multicolumn{2}{c}{Region A} & \multicolumn{2}{c}{Region B} &
\multicolumn{2}{c}{Region C} \\
Run & $\Delta E_\Sigma$ &$\delta E$ & $\Delta E_\Sigma$ &$\delta E$ &
$\Delta E_\Sigma$ &$\delta E$ \\ \hline
I & $0-11.7$ & $3.6-6$ & $0-11.7$ & $0-3.6$ & $14.2-25$ & $1.8-6$ \\
II & $0-8$ & $4.6-6$ & $0-8$ & $0-4.6$ & $13.6-25$ & $1.5-6$ \\
\end{tabular}
\end{ruledtabular}
\end{table}

The values of $\eta_{Si-E}$ and $N_p^{acc}$ are given in
Table~\ref{tab:number} for runs I and II\@.
The total numbers of detected background events,$N_p^{bckg}$, which
belongs to the analyzed region of energies $(\delta E - \Delta
E_\Sigma)$ of protons from reaction (\ref{eq2}a) and met the
criteria (i)--(iii) were defined as
\begin{equation}
  \label{eq:npsigma}
      N_p^{bckg} = N_p^{ff} + N_p^{acc}
\end{equation}
and are also given in Table~\ref{tab:number}.
The uncertainties of $N_p^{bckg}$ include both statistical and
systematical errors.

Based on the measured values $N_p$ and the calculated values
$N_p^{bckg}$ and following Refs.~\cite{helen84,helen83,feldm98}, we
found the yields of detected protons ,$Y_p$, from
reaction~(\ref{eq2}a) in runs I and II\@.
\begin{eqnarray}
\label{eq36b}
      Y_p & = & 7.7^{+4.4}_{-3.4} \qquad \mbox{run I} \nonumber\\
      Y_p & = & 7.5^{+3.8}_{-3.2} \qquad \mbox{run II}
\end{eqnarray}
The errors of $Y_p$ are found in accordance with
Refs.~\cite{helen84,helen83,feldm98} dealing with analysis of small
statistical samples.
In view of Eq.~(\ref{equ:pyieldt}) and the measured values $Y_p$, the
effective rate of nuclear fusion in the $d\mu {}^{3}\mathrm{He}$
complex is obtained from Eq.~(\ref{eq15}).
It can be written as
\begin{equation}
\label{eq37}
      \tilde \lambda_f = \frac{ \lambda_{d\mu} \lambda_\Sigma } {N_\mu
      a_{D/He} W_d\, q_{1s} \, \varphi \, \mathrm{c}_{{}^3\mathrm{He}}
      \lambda_{d {}^3\mathrm{He}}} \frac{Y_p}{\varepsilon_p \,
      \varepsilon_e \, \varepsilon_t \varepsilon_Y }\, ,
\end{equation}
The values of $\tilde \lambda_f$ and $\lambda_\Sigma$ corresponding to
the conditions of runs I and II are given in Table~\ref{tab:10rates}.
Using Eq.~(\ref{equ:efffus}) and the measured effective rates of
nuclear fusion and assuming that $\lambda^1_{f} \ll
\lambda^0_{f}$~\cite{bogda99}, one can get hypothetical estimates of
the partial fusion rate in the $d\mu {}^{3}\mathrm{He}$ complex in its
states with the total orbital momentum $J = 0$
\begin{equation}
\label{eq38}
      \lambda^{J=0}_{f} = \frac{\tilde{\lambda}_{f}(\tilde
      \lambda_{10}+\lambda_{\Sigma}^0 )}{\tilde \lambda_{10}}.
\end{equation}
Table~\ref{tab:10rates} also presents the values for
$\lambda^{J=0}_{f}$ found in runs I and II\@.

\begin{table}[ht]
\begin{ruledtabular}
      \caption{Effective rates of the $1 \to 0$ transition, $\tilde
    \lambda_{10}$, and the nuclear fusion rates in the $d\mu
    {}^{3}\mathrm{He}$ complex.}
\label{tab:10rates}
\begin{tabular}{ccccc}
Run & $\tilde \lambda_{10}$ & $\lambda^{J=0}_{f}$& $\tilde \lambda_f$
& $\lambda_\Sigma$\\
    & [ $10^{11} \, \mbox{s}^{-1}$ ] & [ $10^{5} \, \mbox{s}^{-1}$ ]
& [ $10^5 \, \mbox{s}^{-1}$] & [ $10^{11} \, \mbox{s}^{-1}$ ] \\ \hline
I  & 5.2 & $9.7^{+5.7}_{-2.6}$  & $4.5^{+2.6}_{-2.0}$ & 6.54 \\
II & 7.5 & $12.4^{+6.5}_{-5.4}$ & $6.9^{+3.6}_{-3.0}$ & 6.44 \\
\end{tabular}
\end{ruledtabular}
\end{table}

The averages $\lambda_{\Sigma}^0 = 6 \times 10^{11} \, \mbox{s}^{-1}$
and $ \lambda_{\Sigma}^1 = 7 \times 10^{11} \, \mbox{s}^{-1}$
(averaging over the
data~\cite{kinox93,gersh93,czapl97,belya95c,czapl96c,belya97,czapl96})
were used to get the values presented in Table~\ref{tab:10rates}.
As to the effective rate for transition of the $d\mu
{}^{3}\mathrm{He}$ complex from the state with the angular momentum $J
= 1$ to the state with $J = 0$, it was calculated with allowance for
the entire complicated branched chain of processes accompanying and
competing with the rotational $1-0$ transition (see
Table~\ref{tab:10rates}).
The chain of these processes is considered in detail in
Refs.~\cite{bogda98,bystr99c,bystr99d,bystr99b}.
The effective rates of nuclear fusion in the $d\mu {}^{3}\mathrm{He}$
complex found by us in runs I and II coincide within the measurement
errors.
This is also true for the $d {}^3 \mathrm{He}$ fusion rates
$\lambda^{J=0}_{f}$ obtained by Eq.~(\ref{eq38}).
A comparison of the measured $\lambda^{J=0}_{f}$ with the theoretical
calculations show rather good agreement with \cite{czapl96}, a slight
discrepancy with Refs.~\cite{penko97,bogda99} and considerable
disagreement with Refs.~\cite{nagam89,harle89}.
The cause of this disagreement is not clear yet as also is not clear
the discrepancy between $\lambda^{J=0}_{f}$ calculations in
Refs.~\cite{nagam89,penko97,czapl96,harle89} (see
Table~\ref{tab:rates}).
Note that the theoretical papers
Refs.~\cite{nagam89,penko97,czapl96,harle89,bogda99} yield estimates
with a different degree of approximation.
A correct comparison of the experimental and theoretical
$\lambda^{J=0}_{f}$ is possible only after carrying out some
experiments with the $\mathrm{D}_2 +{}^{3}\mathrm{He}$ mixture ruling
out model dependence on the effective rate of transition of the $d\mu
{}^{3}\mathrm{He}$ complex from the $J = 1$ state to the $J = 0$
state.

A comparison of the results of this paper with the experimental
results~\cite{maevx99} reveals appreciable disagreement between
them.
The shortened form of presentation of the results~\cite{maevx99} does
not allow us to find out sufficiently well the cause of this
considerable disagreement.
Note, however, some results of the intermediate calculations which, to
our mind, disagree with the real estimates of the calculated
quantities.
\renewcommand{\labelenumi}{(\arabic{enumi})}
\begin{enumerate}
\item According to Ref.~\cite{maevx99}, the fraction of the $d\mu$
      atoms which were formed in the excited state under their
      experimental conditions and came to the ground state (per muon
      stop in the target) is $C_{d\mu} = 0.8$.
      The quantity $C_{d\mu}$ is defined as
\begin{equation}
      \label{eq39}
      C_{d\mu} = \frac{1}{2} W_{p,d}(q^{p\mu}_{1s} + q^{d\mu}_{1s}),
\end{equation}
      where $W_{p,d}$ is the probability for direct muon capture by
      the HD molecule followed by formation of the muonic hydrogen
      atom or the excited $d\mu$ atom.
      $q^{p\mu}_{1s}$ and $q^{d\mu}_{1s}$ are the probabilities for
      the transition of the $p\mu$ and $d\mu$ atoms from the excited
      state to the $1s$ ground state.
      According to Refs.~\cite{tresc98c,gartn00,bystr04,tresc99},
      under the Maev~\textit{et.~al} experimental conditions the
      values of the quantities appearing in Eq.~(\ref{eq39}) were
      $W_{p,d} = 0.92$ $q^{p\mu}_{1s} = 0.5$, and $q^{d\mu}_{1s} =
      0.8$.
      Thus, as follows from our estimation, $C_{d\mu} = 0.6$ and not
      0.8 as stated.
\item The number of $d\mu {}^{3}\mathrm{He}$ complexes formed in the
      course of data taking in their experiment was defined as
      \begin{equation}
    \label{eq:39a}
            N_{d\mu {}^{3}\mathrm{He}} = N_\mu C_{d\mu}
            \frac{\lambda_{d\mu {}^{3}\mathrm{He}}}{\lambda_{d\mu}} \,
            ,
      \end{equation}
      and correspond to $N_{d\mu {}^{3}\mathrm{He}} = (4.9\pm 0.4)
      \times 10^8$.

      According to our estimations, the quantities $\lambda_{d\mu
      {}^{3}\mathrm{He}}$, $\lambda_{d\mu}$, $\lambda_{pd\mu}$
      ($pd\mu$ molecule formation rate), and $N_{d\mu
      {}^{3}\mathrm{He}}$ had the values $\lambda_{d\mu
      {}^{3}\mathrm{He}} = 1.32 \times 10^6\ \mathrm{s}^{-1}$
      ($\varphi = 0.0975$ $\mathrm{c}_{{}^3\mathrm{He}}$ = 0.056,
      $\lambda^0_{d^3He} = 2.42 \times 10^8\
      \mathrm{s}^{-1}$)~\cite{bystr04},
      \begin{eqnarray}
      \lambda_{d\mu} & \approx & \lambda_0 + \lambda_ {d\mu
      {}^{3}\mathrm{He}} \varphi \mathrm{c}_{{}^3\mathrm{He}} +
      \lambda_{pd\mu} \varphi \mathrm{c}_p + \tilde \lambda_F \omega_d
      \varphi \mathrm{c}_d \nonumber \\
      & \approx & 2.05\times 10^6\ \mathrm{s}^{-1}
      \end{eqnarray}
      $\lambda_{pd\mu} = 5.6 \times 10^6\, \mathrm{s}^{-1}$, which
      yields $N_{d\mu {}^{3}\mathrm{He}} \approx 3.7 \times 10^8 \,
      \mbox{s}^{-1}$ instead of $(4.9\pm 0.4) \times 10^8 \,
      \mbox{s}^{-1}$.
\item Their ionization chamber detection efficiency for protons from
      reaction (\ref{eq2}a) was defined as $\varepsilon =
      \varepsilon_S \varepsilon_\tau$ and found to be $\varepsilon =
      0.082$, where $\varepsilon_S = 0.13$ is the selection factor for
      events detected in compliance with certain amplitude and
      geometrical criteria, $\varepsilon_\tau = 0.63$ is the time
      factor to take of the fact that the detected events were
      analyzed in the time interval $0.4 \le t \le 1.8 \, \mu$s.
      According to our estimation, $\varepsilon_\tau =
      e^{-\lambda_{d\mu}t_1} -e^{-\lambda_{d\mu}t_2}= 0.44$, because
      under their experimental conditions the $d\mu$ disappearance
      rate is $\lambda_{d\mu} \approx 2.05 \times 10^6 \,
      \mbox{s}^{-1}$, $t_1 = 0.4 \, \mu$s, and $t_2 = 1.8 \, \mu$s.
\end{enumerate}
As can be seen, taking into account only the above items alone the
upper limit of $\tilde \lambda_f$ is, to our mind, appreciably
underestimated in the work of Maev~\textit{et.~al}.
Another cause of this underestimation might be the improper background
subtraction procedure because they determined the background level
using information from earlier experiments~\cite{balin98} carried out
under different conditions and at an experimental facility which was
not completely analogous.
In addition, it is slightly surprising that the background from muon
capture by ${}^{3}\mathrm{He}$ nuclei with the formation of protons in
the energy region near 14.64~MeV is estimated at zero in
Ref.~\cite{maevx99}(see~\footnote{According to Ref.~\cite{bystr04}, the
fraction of protons from muon capture by the ${}^{3}\mathrm{He}$
nucleus in the energy range $14.3-14.64$~MeV per
$\mu{}^{3}\mathrm{He}$ atom is $W^p_{^3He}=2 \times 10^{-6}$.}).

We believe that our $\tilde \lambda_f$ measurement results are
reliable, which is confirmed by stable observation of nuclear fusion
in both runs with the $\mathrm{D}_2 +{}^{3}\mathrm{He}$ mixture
differing in density by a factor of about three.
Nevertheless, as far as the experimental results obtained in this
paper and in Ref.~\cite{maevx99} are concerned, the things are
unfortunately uncertain and need clarifying.

There is a point important for comparison of the calculated
$\lambda^{J=0}_{f}$ with the results of the previous
experiments~\cite{maevx99} and this paper.
Measurement of $\lambda^{J=0}_{f}$ is indirect because it is
determined by Eq.~(\ref{eq38}) with the calculated effective rate for
transition of the $d\mu {}^{3}\mathrm{He}$ complex from the $J = 1$
state to the $J = 0$ state.
Therefore, $\lambda^{J=0}_{f}$ is not uniquely defined and greatly
depends on $\tilde \lambda_{10}$, which in turn is determined by the
chain of processes accompanying and competing with the $1-0$
transition of the $d\mu {}^{3}\mathrm{He}$ complex.
To rule out this lack of uniqueness in determination of
$\lambda^{J=0}_{f}$ and, in addition, to gain information on the
effective $1-0$ transition rate $\tilde \lambda_{10}$ and the nuclear
fusion rate $\lambda^{J=1}_{f}$ in the $d\mu {}^{3}\mathrm{He}$
complex in the $J = 1$ state, it is necessary, as proposed in
Refs.~\cite{bystr99c,bystr99d,bystr99b}, to carry out an experiment
with the $\mathrm{D}_2 +{}^{3}\mathrm{He}$ mixture at least at three
densities in the range $\varphi = 0.03 - 0.2$, where not only protons
from reaction (\ref{eq2}a) but also 6.85~keV $\gamma$~rays should be
analyzed.
Analysis of the results reported in this paper and in
Ref.~\cite{maevx99} makes it possible to put forward some already
obvious proposals as to getting unambiguous and precise information on
important characteristics of $\mu$-molecular ($\lambda_{d\mu
{}^{3}\mathrm{He}}$, $\tilde \lambda_{10}$) and nuclear ($\tilde
\lambda_f$, $\lambda^{J=0}_{f}$, $\lambda^{J=1}_{f}$) processes
occurring in the $\mathrm{D}_2 +{}^{3}\mathrm{He}$ mixture.
It is necessary to conduct experiments at no less than three densities
of the ($\mathrm{HD} + {}^{3}\mathrm{He}$) or ($\mathrm{H}_2 +
\mathrm{D}_3 (1\%) +{}^{3}\mathrm{He}$) mixture with detection of both
protons from reaction (\ref{eq2}a) and 6.85~keV $\gamma$~rays, to
increase at least three times the detection efficiency for protons
$\varepsilon_p$ and for muon decay electrons $\varepsilon_e$ in
comparison with the corresponding efficiencies in the present
experiment.


\begin{acknowledgments}

The authors would like to thanks R.~Jacot-Guillarmod for his help
during the conception of this experiment.
We are thankful to V.F.~Boreiko, A.~Del~Rosso, O.~Huot, V.N.~Pavlov,
V.G.~Sandukovsky, F.M.~Penkov, C.~Petitjean, L.A.~Schaller and
H.~Schneuwly for they help during the construction of the experiments,
the data taking period, and for very useful discussions.
This work was supported by the Russian Foundation for Basic Research,
Grant No.~01--02--16483, the Polish State Committee for Scientific
Research, the Swiss National Science Foundation, and the Paul Scherrer
Institute.

\end{acknowledgments}


\end{document}